\documentclass[11pt,a4paper]{article}
\usepackage{jheppub} 
\usepackage{nameref}
\usepackage{dsfont}
\usepackage{mathrsfs, amssymb, amsmath, amsfonts, latexsym, graphicx, mathabx}  
\usepackage[utf8]{inputenc}
\usepackage{soul}
\usepackage{physics}
\usepackage{accents}
\usepackage[T1]{fontenc}
\usepackage{makecell}
\usepackage{tabularx}
\usepackage{makecell}
\usepackage{array}
\usepackage{pict2e}
\usepackage{float}
\usepackage{diagbox}
\usepackage{hyperref}
\usepackage{tikz-cd}
\usetikzlibrary{arrows.meta,calc}
\usepackage[normalem]{ulem}

\newcommand{\defeq}{\coloneq}

\makeatletter

\DeclareRobustCommand{\loplus}{%
  \mathbin{\mathpalette\dog@lsemi{+}}%
}

\newcommand{\dog@lsemi}[2]{%
  \dog@semi{#1}{#2}{270,90}%
}

\newcommand{\dog@semi}[3]{%
  \begingroup
  \sbox\z@{$\m@th#1#2$}%
  \setlength{\unitlength}{\dimexpr\ht\z@+\dp\z@\relax}%
  \makebox[\wd\z@]{%
    \raisebox{-\dp\z@}{%
      \begin{picture}(1,1)
        \linethickness{\variable@rule{#1}}
        \roundcap
        \put(0.5,0.5){%
          \makebox(0,0){%
            \raisebox{\dp\z@}{$\m@th#1#2$}%
          }%
        }
        \put(0.5,0.5){\arc[#3]{0.5}}
      \end{picture}%
    }%
  }%
  \endgroup
}

\newcommand{\variable@rule}[1]{%
  \fontdimen8
  \ifx#1\displaystyle\textfont3
  \else\ifx#1\textstyle\textfont3
  \else\ifx#1\scriptstyle\scriptfont3
  \else\scriptscriptfont3
  \fi\fi\fi
}

\makeatother

\newcommand{\CC}{\mathbb{C}}
\newcommand{\RR}{\mathbb{R}}
\newcommand{\ZZ}{\mathbb{Z}}
\newcommand{\NN}{\mathbb{N}}

\newcommand{\HH}{\mathbb{H}}

\newcommand{\g}{\mathfrak{g}}

\newcommand{\Diff}{\mathrm{Diff}}
\newcommand{\Vir}{\mathrm{Vir}}
\newcommand{\GL}[1]{\mathrm{GL}(#1,\RR)}
\newcommand{\SL}[1]{\mathrm{SL}(#1,\RR)}

\newcommand{\ISO}[1]{\mathrm{ISO}(#1)}
\newcommand{\U}[1]{\mathrm{U}(#1)}

\newcommand{\UCS}{\mathrm{UCS}}
\newcommand{\ECS}{\mathrm{ECS}}

\newcommand{\QCS}{\mathrm{QCS}}
\newcommand{\BMS}{\mathrm{BMS}}

\newcommand{\Aut}[1]{\mathrm{Aut}(#1)}
\newcommand{\End}[1]{\mathrm{End}(#1)}

\newcommand{\diff}{\mathfrak{diff}}
\newcommand{\gl}[1]{\mathfrak{gl}(#1,\RR)}
\newcommand{\spl}[1]{\mathfrak{sl}(#1,\RR)}

\newcommand{\ecs}{\mathfrak{ecs}}

\newcommand{\qcs}{\mathfrak{qcs}}

\newcommand{\updown}[2]{^{#1}_{\phantom{#1}#2}}

\newcommand{\mf}{\mathfrak}

\newcommand{\mc}{\mathcal}

\newcommand{\weight}[1]{\mathcal E[#1]}

\newcommand{\sect}[1]{\mathbf{#1}}

\newcommand{\overcirc}[1]{\accentset{\circ}{#1}}
\newcommand{\x}{\varphi} 
\newcommand{\zv}[1]{z_{#1}}
\newcommand{\zkm}[1]{w_{#1}}
\newcommand{\V}{\mathrm{V}}
\newcommand{\KM}{\mathrm{KM}}

\newcommand{\ee}{\end{equation}}
\newcommand{\bea}{\begin{eqnarray}}
\newcommand{\eea}{\end{eqnarray}}

\newenvironment{Align}
  {\begin{equation}
   \begin{aligned}}
  {\end{aligned}
   \end{equation}}

\newenvironment{Align*}
  {\begin{equation*}
   \begin{aligned}}
  {\end{aligned}
   \end{equation*}}

\newcommand{\pair}[2]{\langle #1 , #2 \rangle}

\newcommand{\Ad}[1]{\mathrm{Ad}_{#1}}
\newcommand{\coAd}[1]{\mathrm{Ad}^*_{#1}}
\newcommand{\ad}[1]{\mathrm{ad}_{#1}}
\newcommand{\coad}[1]{\mathrm{ad}^*_{#1}}

\newcommand{\lga}{\rhd}

\newcommand{\COC}{\mathsf{C}}
\newcommand{\coc}{\mathsf{c}}
\newcommand{\sCOC}{\mathsf{S}}

\newcommand{\ldv}[1]{\mathcal{L}_{#1}}

\newcommand{\mdot}{\mathbin{\vcenter{\hbox{\scalebox{0.7}{$\bullet$}}}}}

\title{Representation Theory of Three-dimensional Corner Symmetries}
\author[a,b,c]{Giulio Neri,}
\author[d]{Ludovic Varrin}

\affiliation[a]{International School for Advanced Studies (SISSA), \\
Via Bonomea 265, 34136 Trieste, Italy}
\affiliation[b]{National Institute for Nuclear Physics (INFN), \\
Sezione di Trieste, Via Valerio 2, 34127, Italy}
\affiliation[c]{Institute for Fundamental Physics of the Universe (IFPU), \\
Via Beirut 2, 34014 Trieste, Italy}
\affiliation[d]{National Centre for Nuclear Research, Pasteura 7, 02-093 Warsaw, Poland}

\emailAdd{gneri@sissa.it}
\emailAdd{ludovic.varrin@ncbj.gov.pl}

\begin{document}

\abstract{We study the representation theory of the three-dimensional extended corner symmetry group associated with a circular corner. Using induced representation techniques, we construct families of representations of the infinite-dimensional corner symmetry group and analyze their unitarity and irreducibility. We also determine the maximal central extension of the symmetry group and construct the corresponding projective representations. The analysis is motivated by the role that corner symmetries play in quantum gravity, where boundary degrees of freedom --- acted upon by the corner transformations --- are expected to encode aspects of the quantum structure of spacetime. In three dimensions, the representation-theoretic structures that emerge are closely related to symmetry groups and quantum theories that already appeared as central to the study of gravity. This work provides a first step toward a systematic description of quantum gravitational corner degrees of freedom in three dimensions.}

\maketitle
\newpage

\section{Introduction}

Whenever a theory is formulated on a spacetime region with boundary, some of the transformations which would naively be considered as gauge, i.e.~redundancies, become physical. The presence of the boundary ``breaks'' the symmetry, producing a non-zero value for the generators of the symmetry. More explicitly, Noether's second theorem (in its widely used formulation), states that the conserved current associated with a local invariance can be written as the sum of a constrain term, which vanishes on-shell, and a total derivative. Therefore, integrating the current on a spatial slice, the charge localizes on the boundary.
In gravity, where diffeomorphism invariance is a local symmetry, the consequences of this observation have yet to be understood in their full extent. Any codimension-$2$ surface, henceforth called \textit{corner}, that bounds a spatial region can be associated with the following Noether charges
\begin{equation}
\label{eq: Noether charges}
    Q_\xi=\frac{1}{16\pi G_N}\int_S \star \dd \xi^\flat\,,
\end{equation}
which is the index-free expression of the usual Komar charge.
The systematic study of these \textit{corner charges} has attracted a lot of efforts and led to many interesting results in our understanding of classical gravity. In fact, the existence of symmetries imposes restrictions on the allowed configurations and on how the system can transition from one to another, even without explicit knowledge of the form of those solutions.
At the quantum level, the \emph{corner proposal} suggests that focusing on the representation of these charges and their algebra, rather than identifying the appropriate classical phase space to quantize, may help address a number of long-standing questions in quantum gravity—such as the microscopic origin of the Bekenstein--Hawking entropy—in a way that is largely independent of the particular approach to quantum gravity. We review this proposal in Section~\ref{sec: corner proposal}.

The major issues one face when dealing with diffeomorphism-invariance are the \textit{integrability} and the (\textit{non-})\textit{conservation} of the charges. The former arises because the Noether charges~\eqref{eq: Noether charges} cannot be generically associated with the vector field representing the diffeomorphism on the space of fields. Indeed, for some diffeomorphisms, such associated vector field does not even preserve the phase space. The latter problem, separate but related, is due to the open nature of gravitational systems. Even considering the entire spacetime, radiation can still leak out. This issue is very concrete, as it contains the physics of gravitational waves.

In~\cite{ciambelli_embeddings_2022-1}, and previously in~\cite{donnelly_local_2016}, it is shown that the phase space of gravity can be suitably extended by the introduction of \textit{embeddings} in such a way that all diffeomorphisms are realized by Hamiltonian vector fields, thus solving integrability. Moreover, the Hamiltonian function is exactly given by the Noether charge~\eqref{eq: Noether charges}.
We can then immediately see that most diffeomorphisms act trivially on the phase space. If the associated (spacetime) vector field vanishes on $S$ and has zero derivative there, then $Q_\xi=0$. The physical phase space is obtained by removing these degenerate directions. The diffeomorphisms which correspond to a non-zero value of the charge can be classified into:
\begin{itemize}
    \item \textbf{Corner translations} move the corner in the two normal directions. These are not integrable in the non-extended phase space of Einstein--Hilbert gravity;
    \item \textbf{Corner special transformations} correspond to local (i.e.~point-dependent) rotations and boosts of the corner normal plane;
    \item \textbf{Corner diffeomorphisms} are reparametrizations of the corner that move its points tangent to itself.
\end{itemize}
The associated charges form an algebra, which, according to the representation theorem, is isomorphic to a subalgebra of the vector fields on spacetime, up to a central extension. The explicit computation with the \textit{extended phase space} Poisson brackets reveals that there is no central extension at the classical level~\cite{ciambelli_universal_2023}:
\begin{equation}
    \{Q_\xi,Q_\zeta\}=Q_{[\xi,\zeta]}\,.
\end{equation}
For the extended Einstein--Hilbert theory, the charge algebra realize the algebra of the \textit{extended corner symmetry} group
\begin{equation}
\label{eq: ECS definition}
    \ECS:=\Diff(S)\ltimes (\SL{2}\ltimes \RR^2)^S\,.
\end{equation}
The notation $G^S$ denotes the group of functions from $S\to G$, with point-wise multiplication as composition. 

The representation theory of the two-dimensional extended corner symmetry $\ECS_2$, where the corner $S$ is just a point, was addressed in~\cite{ciambelli_quantum_2024,varrin_physical_2024, Neri:2025fsh} and has lead to interesting results~\cite{Varrin:2025okc, Kowalski-Glikman:2025arealaw, ciambelli_entanglement_2026-2}. The present work is the first in the natural extension of those results from 2d to 3d.
In three spacetime dimensions, the corner is one-dimensional and we will assume it has the topology of a circle.
Despite classical gravity being topological in 3d, the topic is highly non-trivial, as demonstrated by the attraction it has had on the scientific community over the course of the last forty years~\cite{Brown:1986nw, Witten:1988hc, Turaev:1992hq, Boulatov:1992vp, Banados:1992gq, Banados:1992wn, banados_three-dimensional_1999, Strominger:1997eq, Carlip:1998wz, Maldacena:1998bw,Witten:2007kt, Maloney:2007ud, Meusburger:2008bs,Barnich:2006av, Barnich:2010eb, Compere:2013bya, asante_holographic_2019-3, goeller_non-perturbative_2020, AndradeeSilva:2023cci, AndradeeSilva:2023okr, Collier:2023fwi, Collier:2024mgv}.

\paragraph{Outline of the paper}
In the remainder of this section, we first summarize the main results of our work. We then briefly review the corner proposal for quantum gravity, which provides our main theoretical motivation, and finally introduce the relevant machinery of Mackey theory to make the paper self-contained.
We use Section~\ref{sec: warmup} as an opportunity to set some notation and as warm-up exercise to familiarize with the application of the little group method and the induction of representations in more simple cases: namely, the two-dimensional corner symmetry group and the continuous-spin representations of the Poincaré group. Section~\ref{sec: 3D case} contains the core of the paper. There, after we describe the group $\ECS_3$ and its properties, we perform the construction of its (non-projective) irreducible representations. We then discuss the unitarity and the irreducibility of these representations in~\ref{sec: measure} and~\ref{sec: irreduibility}. In Section~\ref{sec: corner symmetry central extension} we first determine the maximal central extension of the group $\ECS_3$, which we dub $\QCS_3$. Then, we perform the same construction to determine its standard representations, which are associated with projective representations of the symmetry group.
We conclude in Section~\ref{sec: end} with a recap of the results and an outlook on the relation between the representations of the $\ECS_3$ to other structures that have appeared elsewhere in the study of (quantum) gravity.

\subsection{Summary of the results}

The main result of this paper is the construction of the induced representations of the group $\ECS_3$ associated with a circular corner $S$.
\begin{equation}
    \ECS_3:=\Diff(S)\ltimes (L\SL{2}\ltimes L\RR^2)
\end{equation}
where we introduced the loop group $LG$ notation instead of the more general $G^S$. The normal subgroup $L\RR^2$, despite infinite-dimensional, remains abelian and thus makes viable the application of the little group method devised by Mackey, which we review in Section~\ref{sec: Mackey}.

The dual space to translations is given by \textit{momentum densities}
\begin{equation}
    p=p(\x)\dd \x\,.
\end{equation}
Their orbits under the action of the rest of the group $H:=\Diff(S)\ltimes L\SL{2}$ are quite complicated to characterize fully. Nevertheless, when restricting to nowhere vanishing densities, these orbits become a straightforward \textit{per-point} generalization of the two-dimensional ones. Namely, they are isomorphic to the loop punctured plane (or loop cylinder) $L\RR^2_\circ\equiv L(\RR^2/\{0,0\})$.
For these orbits, the little group contains the loop version of the two-dimensional case, which is the group of 2d shearings $N:=\{\smqty(1&x\\0&1)\mid x\in \RR\}$, but it also includes the action of $\Diff(S)$
\begin{equation}
    H_{p_\star}=\Diff(S)\ltimes LN\,.
\end{equation}
where we picked the constant momentum density $(1,0)\dd\x$ as the orbit representative.

The key realization is that the little group $H_{p_\star}$ is itself a semidirect product with an abelian normal subgroup, thus the study of its representations is amenable to the same treatment.
We find that the shearing parameter $x$ is a density of weight $-2$ on the circle. This brings the group $H_{p_\star}$ close to the $\BMS{}$ group in 3d, whose normal subgroup consisting of supertranslations is an abelian group of densities of weight $-1$. The dual of $LN$ comprises weight-$3$ densities
\begin{equation}
    \nu=\nu(\x)(\dd\x)^3.
\end{equation}
Whereas the momentum density $p$ has a transparent physical interpretation, that of $\nu$ requires working out the induced representations.
Restricting once again to nowhere vanishing densities with constant representative $\nu_\star$ yields the little group $\U{1}\subset\Diff(S)$. The orbits characterized by a single invariant
\begin{equation}
    M:=\oint \nu^{1/3}\in\RR
\end{equation}
where $\oint$ indicates integration over $S$. The existence of this invariant is among the most important results of our analysis. To construct the induced representation, we pick a unitary representation of the little group $\U{1}$, labeled by a \textit{spin} $m\in\ZZ$.

The explicit expression of the induced representations can be given in terms of wavefunctions over the product of the $p$ and $\nu$ orbits, which is isomorphic to $L\RR^2_\circ\times \Diff(S)/S$. The element $(f,h,\alpha)$ of $\Diff(S)\ltimes(L\SL{2}\ltimes L\RR^2)$ is thus represented by
\begin{equation}
    \qty[U_{M,m}{(f,h,\alpha)}\Psi](p,\nu)=\exp\qty[i\oint \pair{p}{\alpha}] \mathbb{D}(f,h;p,\nu)\Psi((f,h)^{-1}\lga p,f^{-1}\lga \nu)\,,
\end{equation}
where $\mathbb{D}$ is the Wigner factor 
\begin{equation}
     \mathbb{D}(f,h;p,\nu)=\exp\qty[im \beta_W(f;\nu)+i\oint \pair{\nu}{x_W(h;p)}],
\end{equation}
with $x_W$ and $\beta_W$ given respectively in~\eqref{eq: Wigner shearing parameter} and~\eqref{eq: shifted Casimir representative}). The pairings always involve densities of total weight $1$.
These representation can be made unitary and irreducible with a proper choice of measure. Indeed, even though Mackey's theorem does not hold for infinite-dimensional groups such as $\ECS_3$ (non-locally compact), we argue for these properties in~\ref{sec: measure} and~\ref{sec: irreduibility}.

The representations $U_{m,M}$ are a new result of our work, but they form only a small subset of what we think are the physically relevant representations. Since quantum mechanics is defined on the projective Hilbert space, where states differing by a phase are identified, we can allows $\ECS_3$ to be represented projectively. The famous result of Bargmann~\cite{Bargmann1954}, extended to groups such as $\ECS_3$ by~\cite{Janssens_2019},\footnote{The key condition is that the group is modeled on a locally convex Lie algebra.} \textit{smooth} projective representations of $\ECS_3^+$ (the component connected to the identity) are one-to-one with \textit{smooth} standard representations of a central extension of it. We determine the maximal central extension of $\ECS^+_3$ as
\begin{equation}
    \QCS_3:=\Vir\ltimes(\widehat{\SL{2}}\ltimes L\RR^2)\,,
\end{equation}
where $\Vir:=\widehat{\Diff}^{+}\!(S)$ is the Virasoro group and $\widehat{\SL{2}}$ is the maximal central extension of $\widetilde{L_e\SL{2}}$ associated with the Kac--Moody algebra extension.

Notably, there is no central extension in the $L\RR^2$ sector. This suggests that the central extension of the normal translations found in~\cite{ciambelli_quantum_2024} may be an artifact of the two-dimensional case. This possibility was already considered there, as quantization does not necessarily commute with dimensional reduction.

The full set of quantum numbers that label a representation of the $\QCS_3$ is thus
\begin{equation}
    \{M,m,\omega,\underline{k}\},
\end{equation}
where $M$ and $m$ were already introduced in the non-projective representations of $\ECS_3$, while $\omega$ is the central charge of the Virasoro extension and $\underline{k}$ is the Kac--Moody level.

\subsection{Corner Proposal}
\label{sec: corner proposal}

The corner symmetry program originates from the observation that gauge theories, and gravity in particular, are intimately characterized by algebras of charges defined on codimension-two surfaces, the \textit{corners}. The possibility of realizing these algebras on a classical phase space is an active area of research, given the potential implications for the quantization of gravity. Unlike bulk gauge symmetries, which annihilate physical states, corner symmetries act non-trivially on the phase space and thus provide an organizing principle for the possible states. A symmetry group that acts canonically and transitively on the phase space provides the basis for Isham quantization~\cite{Isham:1983zr, ISHAM1989349}, which offers one route to realizing the classical symmetry algebra at the quantum level, in the spirit of Dirac's quantization program.

The so-called \textit{corner proposal} reverses this logic: rather than looking for a classical phase space on which a certain symmetry group is realized, the goal is to identify the fundamental corner symmetry group and take it as the starting point. The corner is therefore seen from a group-theoretical perspective as the base on which the representations of the symmetry group are constructed, without requiring an \textit{a priori} embedding into a classical spacetime.
From an ontological perspective, this approach is motivated by the view that the Universe is fundamentally quantum, while the classical world as we see it emerges only in an appropriate limit of a restricted sector of the underlying theory. From this standpoint, it is natural to seek a representation of the corner symmetry group at the quantum level first, and recover the corresponding classical structures from that structure instead of the converse.

The motivation for this proposal is that Einstein gravity is likely an effective theory. Regardless of which fundamental theory ultimately provides its completion, one expects a \textit{mesoscopic} regime in which the fundamental degrees of freedom organize into representations of the corner symmetry group, since we observe the effects of these symmetries at the classical level.
According to Wigner's theorem, the physical states of the gravitational field in a subregion can then be organized by the unitary irreducible representations of the corner symmetry group, and quantities such as entropy, area, and energy are expected to be encoded in representation-theoretic data: characters, weights of coadjoint orbits, and Casimir invariants. Reframing the problem of quantizing gravity in a subregion as the problem of classifying the unitary irreducible representations of a symmetry group places the corner proposal in direct analogy with Wigner's classification of the unitary irreducible representations of the Poincaré group, which laid the foundation for the formulation of relativistic quantum field theory.

In~\cite{ciambelli_universal_2023}, the authors identify the largest possible corner symmetry \textit{algebra}, in the sense that every proper subalgebra of the vector field algebra $\diff(M)$ is also a subalgebra of it. They refer to it as the \textit{universal corner symmetry} algebra, owing to the generality of their derivation. Switching back to group notation, it is given by
\begin{equation}
    \UCS:=\Diff(S)\ltimes (\GL{2}\ltimes \RR^2)^S\,.
\end{equation}
The diffeomorphisms that comprise the $\UCS$ are corner translations $(\RR^2)^S$, general corner linear transformations $(\GL{2})^S$, and corner diffeomorphisms $\Diff(S)$. The extended corner symmetry group~\eqref{eq: ECS definition} we study here is obtained by restricting $\GL{2}$ to $\SL{2}$. This restriction is motivated by the charges realizing the corner symmetry algebra in the extended phase of Einstein-Hilbert gravity
\begin{equation}
Q_\xi=0\,,\qquad \forall\,\xi\in\gl{2}^S\;\,\text{with}\;\,\xi\notin\spl{2}^S\,.
\end{equation}
At present, there is no classical phase space in which the determinant component of $\GL{2}$ is realized non-trivially. Nevertheless, studying the representation theory of $\ECS$ rather than that of $\UCS$ remains a choice, albeit a motivated one.

In this work we take the first step toward a rigorous implementation of the corner proposal, so stated, in the case of three-dimensional gravity. In particular, taking the 3d extended corner symmetry group as the starting point, we provide a first contribution to the classification of its unitary irreducible representations (both standard and projective ones). Applying the Mackey machine we review below to construct induced representations of $\ECS_3$ is a substantial undertaking: the group is both infinite-dimensional and non-locally compact; this places it outside the scope of most of the standard results of Mackey's theory, so that relevant properties and intermediate results must be established independently.

The construction developed here should be viewed as the representation-theoretic backbone on top of which physically relevant computation can subsequently be set up and performed.

\subsection{Mackey's theory}
\label{sec: Mackey}

We give here a self-contained presentation of the induced representation theory and the little group method, which we collectively denote as Mackey's theory after his seminal work on the subject~\cite{Mackey49, Mackey1952, Mackey1953, mackey_theory_1976}. The theory generalizes the results of Wigner and Bargmann on the Poincaré group~\cite{Wigner:1939cj, Wignerbargman1948} to semidirect product groups.
Given the applications that we have in mind, we consider a group of the form $G=H\ltimes A$, where the normal subgroup $A$ is assumed to be abelian (and locally compact). The fact that the product is \textit{semidirect} implies that every element of $H$ can be associated with an automorphisms of $A$, i.e.~there is a map $\sigma:H\to \Aut{A}$, $h\mapsto \sigma_h$.

The starting point of the construction is the unitary dual $\hat A$, consisting of all unitary irreducible representations of $A$ up to equivalence. For an abelian group, all these representations are one-dimensional ($\CC$) so that $\hat A$ is just the set of its (continuous unitary) characters, also known as its Pontryagin dual
\begin{equation}
    \hat A:=\qty{\chi: A \longrightarrow \U{1}}\,.
\end{equation}
If we further restrict to $A$ being a finite-dimensional vector space, then we can identify $\hat A$ with the dual space $A^*$,\footnote{This will be useful because we will later use the hat notation for central extensions.} the set of real linear functionals
\begin{equation}
\label{eq: regular dual}
    A^*:=\qty{p: A\to \RR \mid \text{$p$ is linear}}\,
\end{equation}
and associate each character to a choice of $p\in A^*$ by $\chi_p(\alpha)=e^{i\pair{p}{\alpha}}$, where $\pair{p}{\alpha}$ denotes the natural pairing between a vector space and its dual. In some cases, we will write an explicit form for the pairing; in others, we will keep it implicit: the context will make unambiguous which pairing we are referring to.

The $\sigma$ action of $H$ on $A$ can be extended to $\hat A$ in the following way: given $h\in H$, we can map each representation $\chi$ to a new unitary irreducible representation $h\cdot\chi$ defined by  
\begin{equation}\label{eq: larger group action on representations}
    h\cdot \chi=\chi\circ\sigma_{h^{-1}}\,.
\end{equation}
If $A$ is a vector space, we can write this explicitly as $h\cdot \chi_p=\chi_{\sigma^*_h p}$, where $\sigma^*$ is the action dual --- dual with respect to the pairing between $A$ and $A^*$ --- to $\sigma$. The set $\hat A$ is thus foliated by $H$-orbits.
If the new representation $h\cdot \chi$ has the same character $\chi$, $h$ is said to belong to the so-called \textit{little group} of $\chi$, that is the subgroup of $H$ which leaves the representation/character invariant
\begin{equation}
    H_{\chi}=\{h\in H| h\cdot\chi=\chi\}.
\end{equation}
Every $h\not\in H_\chi$ defines instead a new representation/character, so that each orbit is isomorphic to the coset space
\begin{equation}\label{eq: standard isomorphism in unitary dual}
    \mc O_\chi\cong H/H_\chi\,.
\end{equation}
Once we choose an orbit $\mc O_{\chi}$, the next step is to choose an irreducible unitary representation $D$ of $H_{\chi}$ (living on some vector space $V$) and tensor it with $\chi$ to obtain a representation of the group $G_{\chi}:=H_{\chi}\ltimes A$
\begin{equation}\label{eq: tensorrepresentation}
\begin{split}
    G_{\chi}:=H_{\chi}\ltimes A&\longrightarrow \End{V},\\
    (n,\alpha)&\longmapsto D(n)\otimes \chi(\alpha).
\end{split}
\end{equation}
We note that the intimate relationship between orbits and irreducible representations extend beyond the case of semidirect products and it constitutes the basis for Kirillov's method of coadjoint orbits.

The last step is to induce a representation of the full group $G$ from the representation~\eqref{eq: tensorrepresentation} of its subgroup we just constructed.\footnote{Let $G$ be a group and $\accentset{\curlywedge}{D}:K\to \End{V}$ a representation of a subgroup $K\subset G$. The representation $\accentset{\curlywedge}{D}$ induces a representation $U$ of the full group over the vector space of $V$-valued functions of the left coset space $G/K$, i.e.~$\Gamma(G/K,V)$.
One can construct it explicitly by choosing a section $\sect{s}:G/K\to G$ and taking
\begin{equation*}
    \qty[U{(g)}\psi](p)=\accentset{\curlywedge}{D}(\sect{s}^{-1}(p)\cdot g \cdot\sect{s}(g^{-1}\lga p))\,\psi(g^{-1}\lga p),\quad\forall p\in G/K\,,
\end{equation*}
where $\lga$ denotes the natural action of $G$ on the coset $G/K$. Different choices of section lead to unitarily equivalent representations.} The induced representation is naturally realized on the space of sections of a vector bundle over $G/G_\chi$ with typical fiber $V$.
We assume now that $A$ is a vector space and consider the following chain of isomorphisms
\begin{equation}\label{eq: quotientbylittlegroupisequalmomentumorbit}
    G/G_\chi=(H\ltimes A)/(H_\chi\ltimes A)\cong H/H_\chi \cong \mc O_\chi\subset A^*\,, 
\end{equation}
which lets us parametrize $G/G_\chi$ by points in the dual space.
The first isomorphism descends from the identification $[h]\mapsto [(h,0)]$, while the second can be made explicit by a choice of section
\begin{equation}
    \sect{h}: \mc O_\chi\longrightarrow H,\quad p\mapsto \sect{h}_p\,,
\end{equation}
such that $\chi_p=\sect{h}_p\cdot \chi$. Throughout the paper, we will use bold-face characters to highlight group elements selected by such a choice of section. In the well-known case of massive representations of the Poincaré group, one calls $\sect{h}_p$ the \textit{standard boost}.
We can therefore write the induced representation explicitly as an operator acting on functions of $A^*$
\begin{equation}\label{eq: induced rep, general theory}
    \qty[U{(h,\alpha)}\psi](p)=\chi_p(\alpha)D(\sect{h}_p^{-1} \cdot h \cdot \sect{h}_{\sigma_{h^{-1}}^{*} p})\psi(\sigma_{h^{-1}}^{*}\,p)\,,
\end{equation}
where $\psi : \mathcal{O}_{\chi} \rightarrow V$ is a function on the orbits valued in the vector space of the little group representation.
This representation has three components: a phase produced by the abelian normal subgroup, a translation in the argument due to the semidirect action of $H$ on $A$, and an endomorphism $D$ which represents the little group action. The argument of $D$ is indeed in the little group
\begin{equation}
    (\sect{h}_p^{-1} \cdot h \cdot \sect{h}_{\sigma_{h^{-1}}^{*} p})\cdot\chi=(\sect{h}_p^{-1} \cdot h)\cdot\chi_{\sigma_{h^{-1}}^{*} p}=\sect{h}_p^{-1}\cdot \chi_{\sigma_h^*\sigma_{h^{-1}}^{*} p}=\sect{h}_p^{-1}\cdot \chi_p=\chi
\end{equation}
and it is often called the \textit{Wigner element}, by analogy with the Wigner rotation that appears in the Poincaré group representations.
Since $D$ is a unitary representation, the induced representation~\eqref{eq: induced rep, general theory} can be made unitary as well: a (quasi-)$H$-invariant measure always exists on the orbit $\mc O_\chi$ for locally compact groups, and the functions $\psi$ can be taken to be $L^2$ with respect to that measure. Moreover, Mackey's result on imprimitivity ensures that $U$ is irreducible as well. If the action of $H$ on $\hat A$ is sufficiently regular, this procedure exhaust all irreducible unitary representations of $G$.

Since our goal is not to classify all irreducible representations of
$\ECS_3$ or $\QCS_3$, we will not attempt to verify the full set of
regularity hypotheses entering the standard Wigner--Mackey classification.
Moreover, the groups considered here are infinite-dimensional and
non-locally compact, so that the usual Mackey theorems cannot be applied
directly. We therefore restrict from the outset to a particular class of
orbits and induced representations, and study the unitarity and
irreducibility of the resulting representations separately.

\section{Warm-up: Examples of Induced Representations}
\label{sec: warmup}

Before applying the techniques we presented to the three-dimensional corner symmetries, we consider two relevant examples which serve to separate and illustrate more clearly some technical aspects. The first is the two-dimensional extended corner symmetry group, instrumental to introduce useful elements and set the notation for later sections. The second is the continuous-spin representation of the Poincaré group. While the former will enter the three-dimensional construction directly, the latter will provides a simplified example of the ``two-step procedure'' that we will use later. The confident reader might jump directly to Section~\ref{sec: 3D case}

\subsection{Two-dimensional corner symmetries}
\label{sec: 2D case}

In two-dimensions, the corner $S$ (codimension $2$) is a collection of point, which means that the $\diff(S)$ part of the algebra drops out. What remains is the algebra of the group
\begin{equation}\label{eq: ECS2}
    \ECS_2:= \SL{2}\ltimes \RR^2\,.
\end{equation}
In~\cite{ciambelli_quantum_2024} this group was shown to admit a central extension in the $\RR^2$ sector. The resulting group, dubbed \textit{quantum corner symmetry} group $\QCS_2=\SL{2}\ltimes \HH_3$, has been extensively studied in~\cite{varrin_physical_2024, Varrin:2025okc, Neri:2025fsh, Kowalski-Glikman:2025arealaw, ciambelli_entanglement_2026-2}. The representation theory of the $\QCS_2$ is fully worked out. On the other hand, as we will show in~\ref{sec: corner symmetry central extension}, this central extension does not survive in higher dimensions. For this reason, the two-dimensional case used as an example in what follows will discuss the non-centrally extended $\ECS_2$. The irreducible representations of this group have been classified in~\cite{varrin_physical_2024}. We briefly review here the construction of the unitary irreducible representations as a warm-up for the three-dimensional case.

First of all, since $\RR^2$ is abelian, its unitary dual coincide with the algebraic dual $\RR^{2*}$. Different representations are labeled by a choice of momentum $p\in \RR^{2*}\cong\RR^{2}$
\begin{equation}
    \chi_p: \RR^{2}\longrightarrow \U{1}, \quad \chi_p(\alpha) = e^{i \pair{p}{\alpha}}\,.
\end{equation}
To study the momentum orbits of $\SL{2}$ in $\RR^{2}$, it will be convenient to introduce the Iwasawa decomposition of $\SL{2}$ (also known as KAN decomposition for obvious reasons). Any element $h \in \SL{2}$ can be written uniquely as
\begin{equation}\label{eq: KAN decomposition}
    \sigma^*_h=\mqty(\cos\theta&&-\sin\theta\\\sin\theta&&\cos\theta)\,\mqty(\rho&&0\\0&&\frac{1}{\rho})\,\mqty(1&&x\\0&&1)\equiv k_\theta a_\rho n_x\,,
\end{equation}
which is the composition (in order of applications, from right to left) of a shearing, a boost, and a rotation.
The parameters $\theta \in \qty[0,2\pi]$, $\rho> 0$ and $x\in\mathbb{R}$ are thus coordinates on the group
\begin{equation}
    (\theta,\rho,x)\mapsto h=k_\theta a_\rho n_x\,.
\end{equation}
It is easy to see that every element $p\in\RR^2$ different from the origin $(0,0)$ can be reached from any other by an $\SL{2}$ group element by first scaling the starting vector to the right length, and then rotating it. In other words, the action of $\SL{2}$ on the punctured plane $\RR^2_{\circ}\equiv \RR^2/\{(0,0)\}$ is transitive.
If we take $(1,0)$ as a reference vector, then
\begin{equation}\label{eq: SL transitive action on R2}
    \mqty(p_0\\p_1) = k_{\Theta} a_{R}\, \mqty(1\\0)=:\sect{h}_p\, \mqty(1\\0)\,,
\end{equation}
where $R$ and $\Theta$ are respectively the length of $p$ and its angle with the $x$-axis~\footnote{The function $\mathrm{atan2}(y,x)$ returns the phase of the complex number $x+i y$, hence it correctly takes into account $\pm\pi$ factors which the 1-argument function $\arctan(y/x)$ would miss.}
\begin{equation}\label{eq: coordinates on the punctured plane}
    R=\sqrt{p_0^2+p_1^2},\qquad \Theta=\mathrm{atan2}(p_1,p_0)\,.
\end{equation}

The above construction shows that there exists only two distinct orbits of the $\SL{2}$ action on $\mathbb{R}^2$: the non-trivial orbit $\mc{O}_{(1,0)}=\RR^2_{\circ}$, and the trivial one, containing only the origin $\mc{O}_{(0,0)}$. Since the latter corresponds to a trivial representation of translations, we focus on the former.
Eq.~\eqref{eq: SL transitive action on R2} also shows that the little group of $(1,0)$ is the $N$ component of the Iwasawa decomposition
\begin{equation}\label{eq: 2D little group}
    \SL{2}_{(1,0)}=\qty{n_x \mid x\in\mathbb{R}}=N \cong \mathbb{R}.
\end{equation}
Using the standard isomorphism, the momentum orbit of $p\in \RR^2_\circ$ is given by
\begin{equation}
    \mc{O}_p \cong \SL{2}\slash \SL{2}_{p} = K A \cong S^1 \times \RR_+\,,
\end{equation}
where $\SL{2}_p$ is conjugated to $\SL{2}_{(1,0)}$. This is compatible with our previous observation as the cylinder $S^1 \times \RR_+$ is diffeomorphic to the punctured plane $\RR^2_\circ$.
Since the little group is abelian, its irreducible unitary representations are fully described by its characters
\begin{equation}\label{eq: 2D little group representation}
    D_{\nu}:\SL{2}_{(1,0)}\longrightarrow \U{1}, \quad D_{\nu}(n_x) = e^{i \nu x}\,,
\end{equation}
with distinct representations labeled by different values of $\nu \in \mathbb{R}$. Each representation~\eqref{eq: 2D little group representation} can be straightforwardly extended to the group $\SL{2}_{(1,0)}\ltimes \RR^2$
\begin{equation}
    \accentset{\curlywedge}{D}_{\nu,p}(n,\alpha):=D_{\nu}(n)\chi_p(\alpha)\,.
\end{equation}

We now construct the $\ECS_2$ induced representations which act on functions of the orbit
\begin{equation}
   \qty(\SL{2}\ltimes \RR^2) / \qty(\SL{2}_{p}\ltimes \RR^2)\cong \SL{2}/\SL{2}_p\cong \mc O_p\,.
\end{equation}
Working in the polar coordinates $(\Theta,R)\in S^1\times \RR_+$ defined by~\eqref{eq: coordinates on the punctured plane}, the representation of a group element $(h,\alpha)\in \ECS_2$ is given by
\begin{equation}\label{ECSinducedrepresentation}
    \qty[U_\nu{(h,\alpha)} \psi](p) = e^{i \pair{p}{\alpha}} D_\nu\qty(\sect{h}^{-1}_p \cdot h \cdot \sect{h}_{\sigma^*_{h^{-1}}p})\psi(\sigma^*_{h^{-1}}p).
\end{equation}
where we recall that $\sect{h}_p=k_{\Theta}a_{R}$ is the $\SL{2}$ group element that brings the representative to the point $p$ (cf.~Eq.~\eqref{eq: SL transitive action on R2}), and $\sigma^*_h$ is the $2\times 2$ representation of $h$ given in~\eqref{eq: KAN decomposition}. As expected from the general construction, the Wigner element $\sect{h}^{-1}_p \cdot h \cdot \sect{h}_{\sigma^*_{h^{-1}}p}$ lies in $\SL{2}_{(1,0)}$, the little group of $(1,0)$:
\begin{equation}
\label{eq: Wigner shearing local}
    \sigma^*_{\sect{h}^{-1}_p \cdot h \cdot \sect{h}_{\sigma^*_{h^{-1}}p}} \mqty(1\\0)=\sigma^{*-1}_{\sect{h}_p}\sigma^*_h\sigma^*_{\sect{h}_{\sigma^*_{h^{-1}}p}}\mqty(1\\0)=\sigma^{*-1}_{\sect{h}_p}\sigma^*_h\sigma^*_{h^{-1}}p=\sigma^{*-1}_{\sect{h}_p}p=\mqty(1\\0)\,.
\end{equation}
This means that we can write the Wigner little group element as a shearing transformation $n_{x_W}$, with a parameter $x_W$ that depends on $h$ and $p$.

If we parametrize the $\SL{2}$ group element $h$ by its Iwasawa coordinates $(\theta,\rho,x)$, and the orbit point $p$ by its polar coordinates $(\Theta,R)$, we can write the form of the representation much more explicitly. Let us introduce
\begin{equation}
    \delta := \Theta - \theta\,, \qquad \lambda := \sqrt{\qty(\rho^{-1}\cos\delta - x\rho \sin\delta)^2 + \rho^2 \sin^2\delta}\,,
\end{equation}
where we notice that $\lambda$ is the ratio between the length of the transformed momentum and the length of the original one, $\lambda=\abs{\sigma^*_{h^{-1}}p}/\abs{p}$.
In term of these variables, the $\ECS_2$ representation becomes
\begin{equation}\label{eq: EC2 representations}
   \qty[U_\nu{(\theta,\rho,x,\alpha)}\psi](\Theta,R) = e^{i R\qty(\alpha^0 \cos\Theta + \alpha^1 \sin\Theta)}e^{i \nu x_W}\psi(\Theta',R'),
\end{equation}
with the Wigner shear
\begin{equation}\label{eq: Wigner shearing parameter}
    x_W(h;p)= \frac{2x \cos2\delta + \bqty{\rho^{-2} - \rho^2(1+x^2)}\sin2\delta}{2R^2\lambda^2}=-\frac{1}{R^2}\pdv{\ln\lambda}{\delta}\,.
\end{equation}
and
\begin{equation}\label{eq: coordinates on the punctured plane after transformation}
    R'=\lambda R,\qquad \Theta'=\mathrm{atan2}(\rho \sin\delta,\rho^{-1}\cos\delta - x \rho \sin\delta)\,.
\end{equation}
Using the transformation~\eqref{eq: coordinates on the punctured plane after transformation} of the momentum, we can straightforwardly check that the standard measure $\dd \mu(\Theta,R) = R\dd R \dd\Theta$ is $\SL{2}$-invariant. The representations~\eqref{eq: EC2 representations} is then defined on the Hilbert space of square integrable functions
\begin{equation}
    \norm{\psi}^2 = \int_0^{2\pi} \dd \Theta\int_0^\infty R \dd R\, \abs{\psi(\Theta,R)}^2 < \infty,
\end{equation}
and is unitary with respect to the scalar product
\begin{equation}
      \braket{\x}{\psi} = \int_0^{2\pi} \dd \Theta\int_0^\infty R \dd R\, \bar{\x}(\Theta,R)\psi(\Theta,R).
\end{equation}

In what follows it will be convenient to work on a unitarily equivalent representation defined by $\psi'(\Theta,s)=e^s\psi(\Theta,e^s)$. The Hilbert space of the wavefunctions $\psi'$ is thus
\begin{equation}
    L^2(S^1\times \RR,\dd\Theta \dd s)\,.
\end{equation}
The explicit representation lets us easily derive the action of the $\spl{2}$ generators as operators
\begin{subequations}\label{eq: 2d SL2R generators}
    \begin{align}
    \hat H &= i \cos{2 \Theta}(\partial_s -1) - i\sin{2\Theta}\partial_\Theta + 2\nu e^{-2s}\sin{2\Theta}\,,\\
    \hat E  &= \frac{i}{2}\sin{2\Theta}(\partial_s-1) - i \sin^2{\Theta}\partial_\Theta - \nu e^{-2s}\cos{2\Theta}\,,\\
    \hat F &= \frac{i}{2} \sin{2\Theta}(\partial_s-1) + i \cos^2{\Theta}\partial_\Theta -\nu e^{-2s}\cos{2\Theta}\,.
\end{align}
\end{subequations}
The generators of translations simply act by multiplication
\begin{equation}\label{eq: 2d polar coordinates}
    \hat P_0 = e^s \cos\Theta, \qquad \hat P_1 = e^s \sin\Theta.
\end{equation}
Choosing the trivial representation of the little group ($\nu = 0$) and applying a Fourier transform, one obtains again the defining representation of $\mathrm{ECS}_2$.

The representations~\eqref{eq: EC2 representations} are characterized by a single parameter $\nu \in \mathbb{R}$, reflecting the existence of a unique Casimir operator. In fact, we can show that $\nu$ is precisely the scalar value that $\mc{C}_{\mathrm{ECS}_2}$ takes on these representations
\begin{equation}
\hat{\mc{C}}_{\mathrm{ECS}}\,\psi(\Theta,s) = \qty(-\hat{E}\hat{P_0}^2 + \hat{F}\hat{P}_1^2 + \hat{H}\hat{P_0}\hat{P_1})\psi(\Theta,s)= \nu\,  \psi(\Theta,s).
\end{equation}

\subsection{Continuous-spin representations}
\label{sec: continuous-spin reps}

\begin{figure}
    \centering
    \includegraphics[width=0.8\linewidth, angle=270]{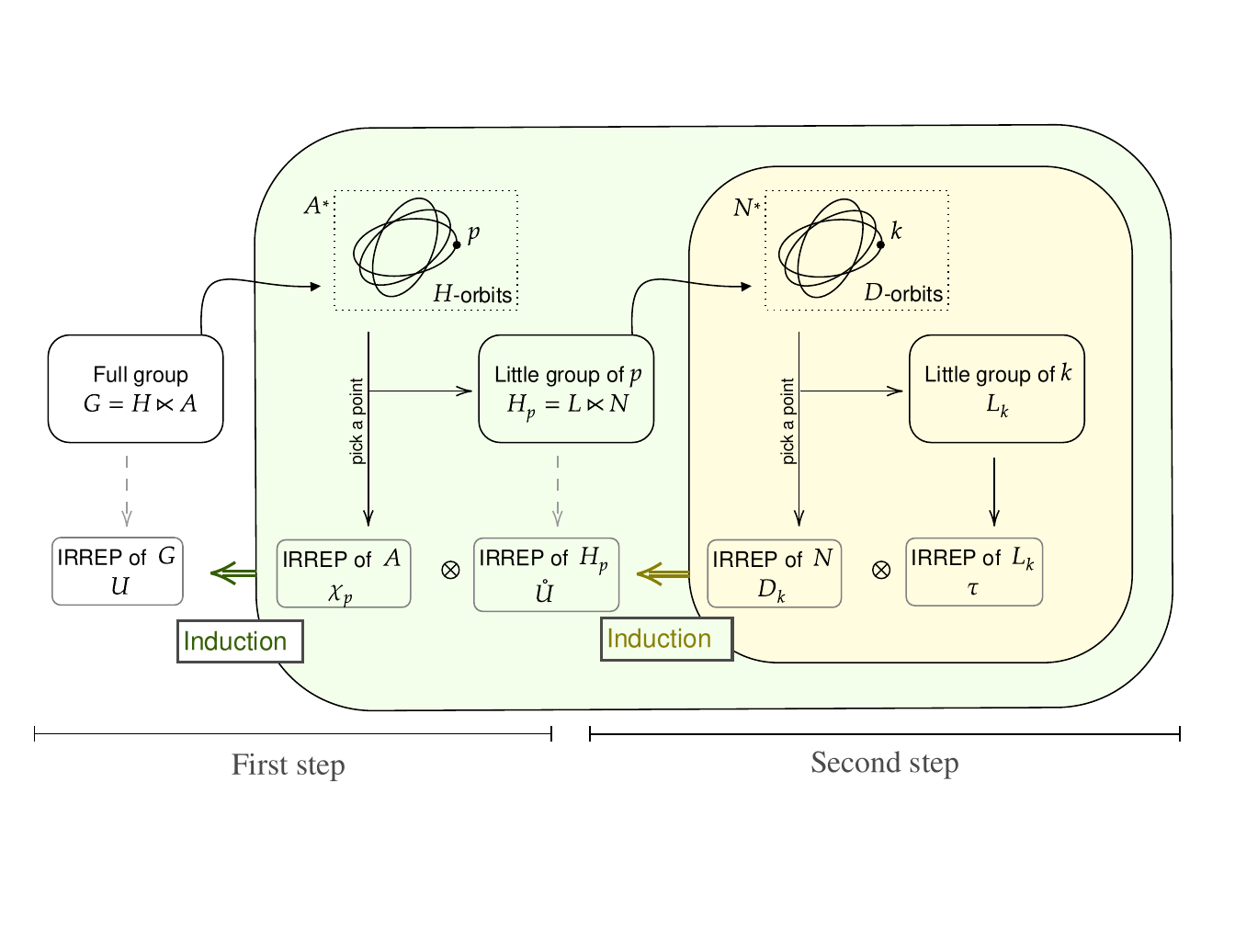}
    \caption{Diagram of the two-step procedure to construct representations of groups that are amenable to Mackey's induction and whose little group is amenable to it as well. We use this procedure in Section~\ref{sec: continuous-spin reps} to find the continuous spin representation of the Poincaré group and in Section~\ref{sec: 3D case} to find the representations of the three-dimensional corner symmetry group.
    The diagram proceeds from left to right. Given the full group $G=H\ltimes A$, of whom one wants to determine the irreducible representations (IRREPs),
    the first step is to study the orbits of $H$ in $A^*$, with $A$ abelian, and pick a representative of the orbit $p$ associated with the character $\chi_p$. In order to construct an IRREP of the little group, which is needed for the induction, one proceeds in the same way because $H_p$ is assumed to have the same form as $G$. This is the second step.
    The construction then proceeds from right to left on the lower row, tensoring representations of the little group with characters and then inducing them to IRREPs of the larger group.}
    \label{fig: two-step induction}
\end{figure}

We now consider another example which will be relevant for the three-dimensional case: the continuous spin representations of the Poincaré group. While these representations are usually deemed unphysical, they provide an example where the little group is also a semidirect product group with abelian normal subgroup, thereby making it amenable to a second application of Mackey's induction. Figure~\ref{fig: two-step induction} describes the general idea of a two-step induction.
Additionally, as all other representations of the Poincaré group, the little group of massless particles is not abelian. The continuous spin representations therefore provide a good warm up exercises before tackling the more complicated three-dimensional corner symmetry group where both of these features will also appear.

We consider the proper orthochronous Poincaré group in four dimensions 
\begin{equation}
  \mathrm{ISO}^+\!(1,3) := \mathrm{SO}^+\!(1,3) \ltimes \RR^{1,3}.
\end{equation}
We denote elements of the Poincaré group by $\qty(\Lambda,x) \in \mathrm{ISO}^+\!(1,3)$.
Translations form an abelian normal subgroup. We denote elements of its dual (both unitary and algebraic)
by $p \in \RR^{1,3*}$ and the pairing with a translation $x \in \RR^{1,3}$ is the standard scalar product $\pair{p}{x} = p_\mu x^\mu$, $\mu= 0,1,2,3$.
The unitary irreducible representations of the abelian subgroup are given by one-dimensional characters
\begin{equation}
    \chi_p: \RR^{1,3}\longrightarrow \U{1}, \quad \chi_p(\alpha) = e^{i p_\mu x^\mu}\,.
\end{equation}
Since the contravariant momentum $p^\mu = \eta^{\mu\nu}p_\nu$ transforms in the fundamental representation
\begin{equation}
  (\Lambda p)^\mu\equiv (\sigma^*_{\Lambda}p)^\mu = \Lambda\updown{\mu}{\nu}p^\nu\,,
\end{equation}
the momentum orbits are famously characterized by the mass hyperbolas 
\begin{equation}
  \mathcal{O}^+_{M} = \qty{p \in \RR^{1,3*} \mid p_\mu p_\nu \eta^{\mu\nu} = M, \, p^0 >0}.
\end{equation}
where the $p^0>0$ condition comes from the fact that we are working with the proper orthochronous Lorentz group.
In the present example, we are interested in massless orbits for which $M=0$. We choose the following contravariant representative
\begin{equation}
  p^\mu_\star = \qty(1, -1,0,0),
\end{equation}
whose little group is given by the two-dimensional Euclidean group
\begin{equation}\label{eq: Euclideanlittlegroup}
  \mathrm{ISO}^+\!(1,3)_{p_\star} = \ISO{2} = \mathrm{SO}(2)\ltimes \RR^2,
\end{equation}
where Euclidean rotations act on the $(2,3)$-plane and Euclidean translations are given by a combination of boosts and rotations in the other two spatial planes.

Following Mackey's construction outlined in the previous section, we would now need to find the irreducible unitary representations of this little group. Since~\eqref{eq: Euclideanlittlegroup} is itself a semidirect product group with abelian normal subgroup, we can construct these representations by aplying Mackey's construction once more. There are therefore two types of massless representations. The first one corresponds to choosing the trivial representations of the Euclidean translations. The stabilizer is then the entire Euclidean rotation group, the representations of which are labeled by an integer $h\in \mathbb{Z}$, the well-known \textit{helicity} of a massless particle. The second type, which is the one we are more interested in, arises from a non-trivial representation of Euclidean translations. These are the so-called continuous spin representations \cite{Wigner:1939cj,Yngvason:1970} (see also \cite{SchusterToro:2013} for a modern field-theory treatment).

The unitary irreducible representations of $\RR^2$ are characterized by elements of its dual $k \in \RR^{2*}$ 
\begin{equation}\label{eq: Euclideancharacters}
    \tau_k: \RR^2 \longrightarrow \U{1}, \quad  \tau_{k}(a) = e^{i \pair{k}{a}}\,.
\end{equation}
where the pairing here is the standard Euclidean product $k_i a^i$, $i=1,2$. It is clear that every non-trivial $k$-orbit is a circle in the dual space
\begin{equation}
    \mathcal{O}_{r} = \qty{(k_1,k_2) \in \RR^{2*} \mid k_1^2 + k_2^2 = r^2, \, r>0}\cong S^1.
\end{equation}
Following the general construction, the induced representation to the Euclidean group are written as acting on functions defined on the circle
\begin{equation}
    \mathcal{H}^r = L^2(S^1,\dd\varphi).
\end{equation}
Since the little group is trivial, applying the general formula~\eqref{eq: induced rep, general theory} gives
\begin{equation}
   \label{eq: continuous spin little group reps} 
   \qty[\overcirc{U}_r{(R_\theta,a)}\psi](\varphi) = e^{ir(a^1 \cos\varphi + a^2 \sin\varphi)}\psi(\varphi-\theta), \quad \psi \in \mathcal{H}^r\,,
\end{equation}
where we used polar coordinates for $k$.
The representation parameter $r>0$ is called the \textit{continuous spin} of the representation.

Now that we have the representations $\overcirc U$ of the little group $\mathrm{ISO}^+\!(1,3)_{p_\star} \subset \mathrm{ISO}(1,3)$, we need to induce it to the rest of the Poincaré group.
In order to construct the Wigner element, we parametrize an element of the orbit $p\in \mathcal{O}^+_{p_\star}$ by
\begin{equation}
    p^\mu = p^0(1,-\cos\theta_p,\sin\theta_p \cos \varphi_p, \sin\theta_p\sin\varphi_p).
\end{equation}
and choose the following section
\begin{equation}
    \sect{\Lambda}_p = R(\vec{u},\theta_p)B_1(\ln p^0),
\end{equation}
where $p^0 >0$, $R(\vec{u},\theta_p)$ is a rotation of angle $\theta_p$ around the axis $\vec{u} = (0,\sin\varphi_p,-\cos\varphi_p)$ and $B_{1}(\eta)$ is a boost along the first axis of rapidity $\eta$,
such that
\begin{equation}
    \sect{\Lambda}_p\, p_\star\equiv \sigma_{\sect{\Lambda}_p}^*  p_\star = p.
\end{equation}
The Wigner element is then defined as
\begin{equation}\label{eq: wignerelementcontinuousspin}
    (\Lambda;p)_W := \sect{\Lambda}^{-1}_p \Lambda \sect{\Lambda}_{\Lambda^{-1}p}.
\end{equation}
Since this must be an element of the little group, the above equation defines parameters $(\theta_W,a_W)\in \mathrm{ISO}^+\!(1,3)_{p_\star}$ that enter the little group representation~\eqref{eq: continuous spin little group reps}. The full continuous spin representation of the Poincaré group can finally be written on wavefunctions $\Psi:\mathcal{O}^{+}_{p_\star}\times S^1 \rightarrow \CC$
\begin{equation}
   \qty[U_r{(\Lambda,x)}\Psi](p,\varphi) = e^{ipx}e^{i r\qty[a^1_W(\Lambda;p)\cos\varphi + a^2_W(\Lambda;p)\sin\varphi]}\Psi(\Lambda^{-1}p,\varphi-\theta_W(\Lambda;p)).
\end{equation}
The (Lorentzian) translations act only via a phase as is expected from Mackey's construction. The parameters $a_W,\theta_W$
are implicitly defined by equation~\eqref{eq: wignerelementcontinuousspin}: in order to compute them explicitly for a given group transformation, one needs to write the Wigner element in the parametrization of the little group used in writing out its representation.

Finally, we note that, if the little group inside of $\mathrm
{ISO}(2)$ was not trivial, there would be an additional choice of section in the first induction~\eqref{eq: continuous spin little group reps}. This will an additional complication for the three-dimensional corner symmetry case that we treat next. As exemplified here, its (irreducible) representations can be constructed in two steps: (1) consider $G=H\ltimes A$ with $A$ an abelian normal subgroup, as in Section~\ref{sec: Mackey}, study the orbits of $H$ in $A^*$ and identify the little group $H_p$ of a certain representation $\chi_p\,$; (2) then write the little group as $H_p=L\ltimes N$, with $N$ an abelian normal subgroup, and apply the same procedure to $H_p$ to determine the little group $L_k$ of a certain representation $D_k$. The induction then proceeds in the other direction, extending the tensor representation $D_k\otimes \tau$ to a representation $H_p$ and then extending the tensor representation $\overcirc U\otimes \chi_p$ to a representation of $G$, as illustrated in Figure~\ref{fig: two-step induction}, where all elements are defined. Mackey's theory guarantees that, if the chosen representations are irreducible, all induced representations so constructed will be irreducible as well.

\section{Corner Symmetries in Three Dimensions}\label{sec: 3D case}

In three dimensions, a corner is a one-dimensional submanifold. In this work, we assume that the corner has the topology of a circle $S\cong S^1$, a \textit{loop}. We can thus write the corner symmetry group as
\begin{equation}\label{eq: ECS3}
    \ECS_3:=\Diff(S)\ltimes L(\SL{2}\ltimes \RR^2)
\end{equation}
where $LG$ is the loop group of $G$, that is the group of smooth functions $g\in C^\infty(S^1,G)$, with composition given by point-wise multiplication
\begin{equation}
    (g_1\cdot g_2)(\x)=g_1(\x)\cdot g_2(\x), \qquad \x\in [0,2\pi)\,,
\end{equation}
See Appendix~\ref{app: loop group theory} for a comprehensive review of the basic theory of loop groups. For consistency, we take $\Diff(S)$ to be the group of smooth diffeomorphisms $\Diff^\infty(S)$. We will return on this regularity when discussing the unitarity and irreducibility of the induced representations.
In our case, we consider functions that are valued in $\SL{2}\ltimes \RR^2$, which is just the two-dimensional corner symmetry group $\ECS_2$. Since, as a set, $\ECS_2$ is just a Cartesian product, we can think of each loop group element $g:\x\to (h(\x),\alpha(\x))$ as a pair of two loop group elements $h:\x\mapsto h(\x)$ and $\alpha:\x\mapsto\alpha(\x)$, which means that
\begin{equation}
    L(\SL{2}\ltimes \RR^2)=L\SL{2}\ltimes L\RR^2.
\end{equation}

The group of circle diffeomorphisms $\Diff(S)$ has two connected components~\cite{oblak_bms_2017}, connected by the inversion $\x\mapsto 2\pi-\x$. For connected groups, the number of connected components of its loop group (the zero-th homotopy group) can be obtained from the general property $\pi_0(LG)\cong \pi_1(G)$.\footnote{The general relation for a connected $G$ is $\pi_k(LG)\cong \pi_k(G)\times \pi_{k+1}(G)$} This means that $\pi_0(L\SL{2})=\pi_1(\SL{2})=\ZZ$. The group $\ECS_3$ has thus two infinite sets of connected components, characterized by whether the diffeomorphism preserves the orientation or not, and by the winding number of $h$.
In the following, we consider only the component connected with the identity
\begin{equation}
    \ECS_3^+:=\Diff^{+}\!(S)\ltimes (L_e\SL{2}\ltimes L\RR^2)\,.
\end{equation}
Notice that $L_e\SL{2}\cong \SL{2}\times L_0 \SL{2}$ where $L_0 G$ denotes the \textit{based} loop group, i.e.~the subgroup of $LG$ such that $g(0)=e$. This means that $\pi_1(L_e\SL{2})\cong \pi_1(\SL{2})\times \pi_2(\SL{2})=\ZZ$. Moreover, since $\pi_1(\Diff^{+}\!(S))=\ZZ$ as well, the first homotopy group of the full group is
\begin{equation}
    \pi_1(\ECS_3^+)=\ZZ\times \ZZ\,,
\end{equation}
where the first $\ZZ$ is the winding in the diffeomorphism group, the second is the winding in $\SL{2}$ --- different from that of $L\SL{2}$: the universal cover of $\ECS_3^+$ is obtained by unwinding these two circle directions.

The algebra of the loop group $LG$ is given in terms of the Lie algebra $\g$ as the set of smooth functions valued in the algebra $L\g\equiv C^\infty(S^1,\g)$. A dense subalgebra is given by direct products of $\g$ with Laurent polynomials $\CC[z,z^{-1}]$, $z=e^{i\x}$ (i.e.~Fourier modes). In the case of $L\spl{2}\ltimes L\RR^2$, these read
\begin{equation}\label{eq: loop algebra generators}
    J^\alpha_n\equiv J^\alpha\otimes z^n, \quad P_n^a\equiv P^a\otimes z^n, \qquad n\in\ZZ\,,
\end{equation}
where $\{J^\alpha,P^a\}$, $\alpha\in\{-,0,+\}$, $a\in\{0,1\}$ form a basis of $\spl{2}\ltimes \RR^2=\ecs_2$.
For convenience, we collectively denote these generators as $V_A$, with a super-index $A=(\alpha,a)$.
The brackets are then inherited from point-wise multiplication
\begin{equation}
    [V_n^A, V_m^B]=[V^A,V^B]\otimes z^{n+m}=C^{AB}{}_C V^C_{n+m},
\end{equation}
where $C^{AB}{}_C$ are the structure constants of the two-dimensional algebra $\mathfrak{ecs}_2$.
The Lie algebra of the group $\Diff(S)$ is the space of vector
fields on the circle, with the Lie bracket given by the opposite of the standard
Lie bracket between vector fields~\cite{oblak_bms_2017}. If we introduce the basis
\begin{equation}
    \ell_m\equiv e^{im\x} \partial_\x\,\qquad m\in\ZZ\,,
\end{equation}
the Lie bracket takes the form
\begin{equation}\label{eq: dewitt brackets}
    [\ell_m, \ell_n] = (m-n)\ell_{m+n}\,,
\end{equation}
which is the standard \textit{de Witt} algebra of conformal field theory.

The diffeomorphisms act on the loop by reparametrization, that is with the Lie derivative of the corresponding vector field. In modes, this reads
\begin{equation}
    [\ell_m, V^A_n]=-n V^A_{m+n}\,.
\end{equation}
Notice that both families of generators, $J_n^\alpha$ and $P_n^a$, transform in the same way under diffeomorphisms: this condition will be the source of obstruction for the central extension discussed in the next section.

\subsection{First step: momentum density orbit}
\label{sec: momentum density orbit}

The starting point of this program is the realization that, despite being infinite-dimensional, the group $L\RR^2$ is still abelian. Its unitary irreducible representations are therefore still just characters, indexed by a loop version of the momentum in $L\RR^{2*}$
\begin{equation}
    \chi_p: L\RR^2\longrightarrow \U{1}, \quad \chi_p(\alpha) = e^{i \oint\pair{p}{\alpha}},\quad p\in L\RR^{2*}\,,
\end{equation}
where we denoted by $\oint$ the integral over the circle.\footnote{By definition, the pairing between a loop group element and one of its dual should include the integral over $S^1$ --- otherwise the result would not be a number --- but we have chosen to slightly abuse our notation in order to separate more clearly the results holding point-wise on the circle from those holding only after integration.}
Since the translation parameter transform as a scalar (that is, it has weight zero, see~\cite{oblak_bms_2017} for a rigorous definition), the only way for $\pair{p}{\alpha}$ to have the right diffeomorphism-covariance properties is for $p$ to transform as a density of weight $1$
\begin{equation}\label{eq: momentum density transformation law}
    p\mapsto f_* p,\quad (f_* p)(\x)=p(f^{-1}(\x))\pdv{f^{-1}(\x)}{\x}\,.
\end{equation}
Introducing the notation $\weight{k}$ for the set of weight-$k$ densities, we can say that $L\RR^2=\weight{0}$, while $L\RR^{2*}=\weight{1}$. More generally, the dual space of $\weight{k}$ is $\weight{1-k}$.
A remark is in order. In contrast to what we said around Eq.~\eqref{eq: regular dual}, the appropriate notion of dual of $L\RR^2$, interpreted as the set of continuous functions $S^1\to\RR^2$, is the continuous linear dual, i.e.~the set of linear functions of $L\RR^2$ that are continuous. Generically, these are distributions. Restricting to distributions which can be written as $\oint\pair{p}{\cdot}$ with a continuous $p:S^1\to \RR^{2*}$ is indeed a \textit{restriction} --- often invoked in the literature with the term \textit{regular} dual~\cite{oblak_bms_2017} --- which we plan to dispense with in a future work.

Since the multiplication of loop group is point-wise, everything that was said in Section~\ref{sec: 2D case} about the action of $\SL{2}$ on $\RR^2$ will hold as well for their loop version. In particular, any nowhere-zero momentum density can always be mapped to the constant momentum density $p_\star=(1,0)\dd\x$ by the smooth $\SL{2}$-valued map $\x\mapsto \sect{h}_{p(\x)}^{-1}$. On the other hand, if the momentum density crosses the origin at some $\x=\x_0$, there is no smooth map which sends it to $p_\star$ for all $\x\ne \x_0$.
Momentum densities with different number of such crossing points all belong to different orbits of $L_e\SL{2}$ on $L\RR^2$, characterized by that number only. As for the orbit of diffeomorphisms, the full range of possibilities is much wider, depending on whether $p(\x)=(0,0)$ at a single point or on an interval.

There are two orbits which most straightforwardly generalize the local picture we gave in Section~\ref{sec: 2D case}, namely
\begin{itemize}
    \item the origin, with the only representative $(0,0)\dd\x$;
    \item the loop punctured plane $L\RR^2_\circ$, with the representative $p_\star=(1,0)\dd\x$.
\end{itemize}
The little group of the first orbit is the group $\Diff^{+}\!(S)\ltimes L_e\SL{2}$. A representations of this group is given by the WZW model for $\SL{2}$~\cite{Maldacena:2000hw}.
As for the second orbit, its little group surely contains a loop version of~\eqref{eq: 2D little group}, that is $LN$. On the other hand, given the transformation law~\eqref{eq: momentum density transformation law}, it is clear that the action of a diffeomorphism on a non-zero constant density such as $p_\star$ is just a scaling. Moreover, for any single vector, a scaling can be compensated with a boost. Explicitly, we can use
\begin{equation}
    \label{eq: stabilizing boost}
    \sigma^*_{a_f(\x)}=\mqty(\bqty{(f^{-1})'(\x)}^{-1}&0\\0&(f^{-1})'(\x))\in LA
\end{equation}
The little group is therefore
\begin{equation}\label{eq: 3D little group}
    H_{p_\star}=\qty{(f,n_x a_f) \mid f\in\Diff^{+}\!(S),\;x\in L\mathbb{R}}\cong\Diff^{+}\!(S)\ltimes LN\,,
\end{equation}
which can be written as the semidirect product of $\Diff^{+}\!(S)$ with the two-dimensional little group~\eqref{eq: 2D little group}. The nature of this semidirect product follows from the composition law of $H_{p_\star}$, which we determine in Appendix~\ref{app: loop group theory}, Eq.~\eqref{eq: little group composition law}:
\begin{equation}
    (f_1, n_{x_1} a_{f_1})\cdot(f_{2}, n_{x_2} a_{f_2})=(f_1\circ f_2, n_{x_{12}} a_{f_1\circ f_2}),
\end{equation}
with
\begin{equation}
    x_{12}(\x)=x_1(\x)+x_2(f_1^{-1}(\x))\pqty{\pdv{f_1^{-1}(\x)}{\x}}^{-2}\,.
\end{equation}
Setting $x_1=0$ and $f_2=\text{id}$ --- and renaming $f_1=f$, $x_2=x$, $x_{12}=x'$ --- this shows that $x\in\weight{-2}$, since it transforms as
\begin{equation}\label{eq: antiquadratic density transformation}
    x'(\x)=x(f^{-1}(\x))\pqty{\pdv{f^{-1}(\x)}{\x}}^{-2}\,.
\end{equation}

To induce a representation of the $\ECS_3$, we now need to pick a representation of the little group $H_{p_\star}\cong \Diff^{+}\!(S)\ltimes \weight{-2}$, which we will later tensor with the representation $\chi_p$.
The group structure of $H_{p_\star}$ bears a strong resemblence to the (extended) three-dimensional BMS group $\Diff^{+}\!(S)\ltimes \diff(S)$, where $\Diff^{+}\!(S)$ is the group of superrotations and $\diff(S)$ --- its Lie algebra --- is the abelian group of supertranslations~\cite{Barnich:2009se, Barnich:2010eb, oblak_bms_2017}. Since vectors are just densities of weight $-1$, we can write the group in a more suggestive form $e\mathrm{BMS}_3^+=\Diff^{+}\!(S)\ltimes \weight{-1}$. Additionally, both $H_{p_\star}$ and $e\mathrm{BMS}_3^+$ can be seen as instances of a more general family $\mathrm{BMS}_w:=\Diff^{+}\!(S)\ltimes \weight{w}$ in which the abelian `supertranslations' are represented by weight-$w$ densities.

\subsection{Second step:  Casimir density orbit}
\label{sec: Casimir density orbit}

The construction of induced representation instructs us to choose a representation of the little group $H_{p_\star}=\Diff^{+}\!(S)\ltimes \weight{-2}$. Since this is yet another semidirect product with an abelian normal subgroup, we can apply Mackey's little group method to it again, following the logic exemplified by the Poincaré continuous spin representations and illustrated in the diagram~\ref{fig: two-step induction}. Therefore, we fix an element in the \textit{regular} dual of $\weight{-2}$, which is $\weight{3}$, and construct the character
\begin{equation}
    D_{\nu}: LN\longrightarrow \U{1}, \quad D_\nu(n_x) = e^{i \oint \pair{\nu}{x}}\,,
\end{equation}
where $\nu$ is a weight-3 density, so that its component transforms as
\begin{equation}\label{eq: cubic density transformation}
    (f\lga \nu)(\x)=\nu(f^{-1}(\x))\pqty{\pdv{f^{-1}(\x)}{\x}}^{3}\,
\end{equation}
under a circle diffeomorphism. From now one, we are going to denote a group action on the dual by $\lga$, consistently with the isomorphism between orbits and group cosets.

Given this transformation law, it is clear that the $\Diff^{+}\!(S)$ orbit in $\weight{3}$ depend on the number of zeros of the function $\nu(\x)$, and, given a certain set of zeros, on the degree with which the function vanishes there.
In the following, we assume that $\nu(\x)$ is either
\begin{itemize}
    \item zero everywhere, whose orbit is trivial;
    \item nowhere vanishing. Unlike the previous case, where $L\SL{2}$ could be used to connect the orbits of all nowhere-vanishing momentum densities, the invariance of the cubic differential $\nu\equiv \nu(\x)(\dd\x)^3$ implies that densities with different values of
    \begin{equation}
     M=\oint \nu^{1/3}\in\RR
    \end{equation}
    belong to different $\Diff^{+}\!(S)$ orbits. On the other hand, every element with the same $M$ can be connected to the constant density $\nu_\star=\frac{M^3}{8\pi^3}(\dd\x)^3$, meaning that each orbit is only labeled by $M$.\footnote{An explicit diffeomorphism that connects $\nu$ to $\nu_\star$ is given by
    \begin{equation}
    \label{eq: map from Casimir to Casimir representative}
    f(\x)=\frac{2\pi}{M}\int^{\x}_0 \nu^{1/3}(\x')\dd\x'.    
    \end{equation}}
\end{itemize}

The stabilizer of the zero density is the full group $\Diff^{+}\!(S)$, whose representation theory has been studied by many authors~\cite{Kirillov1981, Kirillov:1982kav,segal_unitary_1981,witten_coadjoint_1988, alekseev_path_1989, Shavgulidze1978, Shavgulidze1988, Shavgulidze1995, Shavgulidze1997}. We consider here the second case. The little group of the constant representative $\nu_\star$ is given by rigid rotations, as it follows straightforwardly from~\eqref{eq: cubic density transformation}
\begin{equation}
    \pdv{f^{-1}(\x)}{\x}=1\implies f\in \U{1},
\end{equation}
\begin{equation}\label{eq: 3d Casimir little group}
    \U{1} = \qty{r_\beta :S^1 \rightarrow S^1 \mid r_\beta(\x) = \x + \beta \; \mathrm{mod}2\pi}\,.
\end{equation}
It follows from the standard isomorphism that the orbit $\mc O_{\nu_\star}$ is then $\Diff^{+}\!(S)/S$.
The unitary representations of the little group $\U{1}$ are then characterized by an integer $m\in\mathbb{Z}$, which we call the \textit{spin} of the representation
\begin{equation}
    \tau_m: \U{1}\longrightarrow \U{1}, \quad \tau_m(r_\beta) = e^{i m \beta}\,.
\end{equation}

We construct a representation of $\U{1}\ltimes LN$ by tensoring $\tau_m$ and $D_\nu$, and induce from this a representation of the group $\Diff^{+}\!(S)\ltimes LN$, which --- we remind the reader --- is the little group of the constant momentum density (cfr.~Eq.~\eqref{eq: 3D little group}). The induced representation acts on functions of the coset space $\Diff^{+}\!(S)/S$. The group of diffeomorphisms connected to the identity admits the decomposition~\cite{Lurie2009}
\begin{equation}
    \Diff^{+}\!(S)=\Diff^{+}_0\!(S)\circ \U{1}\,,
\end{equation}
where $\Diff^{+}_0\!(S)$ is the group of \textit{based} (orientation-preserving) diffeomorphisms that leave the point $\x=0$ fixed. As manifolds, we can thus  identify the coset space $\Diff^{+}\!(S)/S$ with the $\Diff^{+}_0\!(S)$:
this group is a very natural global section of the principal $S^1$-bundle $\Diff^{+}\!(S)\to\Diff^{+}\!(S)/S$. Given a representative $f$ of an  equivalence class $[f]\in\Diff^{+}\!(S)/S$, the section can be realized explicitly by
\begin{equation}
    f \mapsto f_0:=f\circ r_{f^{-1}(0)}\in\Diff^{+}_0\!(S)\,,
\end{equation}
so that, trivially, $f=f_0\circ r^{-1}_{f^{-1}(0)}$. Since rotations are abelian, it is clear that a different representative $f'$ of the same class would define the same $f_0$. We then choose to parametrize the orbit elements by the section $\nu\to \sect{f}_\nu$, where (cfr.~\eqref{eq: map from Casimir to Casimir representative})
\begin{equation}
    \label{eq: standard based diffeo}\sect{f}^{-1}_\nu(\x)=\frac{2\pi}{M}\int^{\x}_0 \nu^{1/3}(\x')\dd\x'\,.
\end{equation}
Acting on $\nu$ with a diffeomorphism $f^{-1}$ maps $\nu$ to a new element $f^{-1}\lga \nu=(f^{-1}\circ \sect{f}_\nu)\lga \nu_\star $. The diffeomorphism $f^{-1}\circ \sect{f}_\nu$ is generically not in $\Diff^{+}_0\!(S)$, but differs from the chosen representative in the same equivalence class by a rigid rotation. One can check explicitly that
\begin{equation}
\label{eq: shifted Casimir representative}
    \sect{f}_{f^{-1}\lga\nu}^{-1}(\x)=\sect{f}_{\nu}^{-1}(f(\x))-\beta_W\,,\quad\text{with}\;\;\beta_W(f;\nu)=\frac{2\pi}{M}\int_0^{f(0)}\nu^{1/3}(\x)\dd\x\,,
\end{equation}
hence we deduce that $\beta_W$ coincides with the parameter of the Wigner little group element 
\begin{equation}
\label{eq: U(1) little group element}
    W(f;\nu):=\sect{f}_\nu^{-1}\circ f \circ \sect{f}_{f^{-1}\lga\nu}=\sect{f}^{-1}_\nu\circ f\circ (f^{-1}\circ \sect{f}_\nu\circ r_{\beta_W})=r_{\beta_W}\,.
\end{equation}
Denoting an element of $\Diff^{+}\!(S)\ltimes \weight{-2}$ as $(f,x)$, the representation then takes the form
\begin{equation}\label{eq: DiffLN representation}
    \qty[\overcirc{U}_{M,m}{(f,x)}\psi](\nu)=\tau_m(r_{\beta_W(f;\nu)})D_{\nu}(n_x)\psi(f^{-1}\lga\nu)\,,
\end{equation}
where $\psi$ is a wavefunction in $\Gamma(\Diff^{+}\!(S)/S,\CC)$.

The full representation of $\ECS^+_3$ is induced from the tensor product representation $\overcirc{U}_{M,m}\otimes\chi_p$ on wavefunctions of the momentum density, valued in the space of sections $\Gamma(\Diff^{+}\!(S)/S,\CC)$. Given the constant momentum density representative $p_\star=(1,0)\dd \x$, we can parametrize the orbit by
\begin{equation}
\label{eq: momentum orbit parametrization}
    p=(\text{id},\sect{h}_p)\lga p_\star=\sigma^*_{\sect{h}_p}p_\star\,,
\end{equation}
where $\sect{h}_p(\x)$ is given by $\sect{h}_{p(\x)}$ according to~\eqref{eq: SL transitive action on R2}. The reader should now be familiar with the evaluation of the Wigner element associated with $p$ and a $\Diff^{+}\!(S)\ltimes L_e\SL{2}$ transformation.\footnote{Gathering enough patience, we proceed straightforwardly
\begin{equation}
    (\text{id},\sect{h}_p)^{-1}\cdot(f,h)\cdot (\text{id},\sect{h}_{(f,h)^{-1}\lga p})=(f,\sect{h}^{-1}_p\cdot h \cdot f_*\sect{h}_{(f,h)^{-1}\lga p})\,.
\end{equation}
Using the coadjoint action~\eqref{eq: coadjoint action of DiffLSL}, we evaluate the argument of the last $\sect{h}$ on the r.h.s. 
\begin{equation}
    (f,h)^{-1}\lga p=(f^{-1},f^*h^{-1})\lga p=\sigma^*_{f^*h^{-1}}f^* p=f^*(\sigma^*_{h^{-1}} p)\,.
\end{equation}
We then compute
\begin{equation}
    f^*(\sigma^*_{h^{-1}} p)=\sigma^*_{\sect{h}_{f^*(\sigma^*_{h^{-1}} p)}}p_\star=(f^*\sigma^*_{\sect{h}_{\sigma^*_{h^{-1}} p}})p_\star=f^*(\sigma^*_{\sect{h}_{\sigma^*_{h^{-1}} p}}\cdot\sigma^*_{a_f}p_\star)\,,
\end{equation}
which implies that $f_*\sect{h}_{(f,h)^{-1}\lga p}=\sect{h}_{\sigma^*_{h^{-1}} p} a_f$. We can then finally simplify the Wigner element computation as
\begin{equation}
    (f,\sect{h}^{-1}_p\cdot h \cdot f_*\sect{h}_{(f,h)^{-1}\lga p})=(f,\sect{h}^{-1}_p\cdot h \cdot \sect{h}_{\sigma^*_{h^{-1}} p} \cdot a_f)=(f,n_{x_W} a_f)\,,
\end{equation}
where we recognized the (point-wise) definition of the Wigner shearing~\eqref{eq: Wigner shearing local}.
}
Explicitly, the components of the Wigner element $(f,x)_W$ are given by
\begin{equation}
\label{eq: Wigner element for momentum orbits}
    f_W((f,h);p)=f,\qquad x_W((f,h);p)=f\lga x_W(h;p)\,,
\end{equation}
where $x_W$ is the shearing parameter locally given by~\eqref{eq: Wigner shearing parameter}. Given the result in~\eqref{eq: antiquadratic density transformation}, this implies that $x_W\in\weight{-2}$, compatibly with its scaling as $R^{-2}$ --- indeed $p\in\weight{0}$ implies $R\in\weight{1}$ and $\delta\in\weight{0}$. Compared to the loop-version of the two-dimensional case, the little group element includes a diffeomorphism and the associated boost.

We can finally write down the representation of the $\ECS_3^+$ as
\begin{equation}\label{eq: implicit ECS3 representation}
    \biggl[U_{M,m}{(f,h,\alpha)}\Psi\biggr](p)=\chi_p(\alpha)\,\qty[\overcirc{U}_{M,m}{(f,n_{x_W})}\Psi]((f,h)^{-1}\lga p)\,,
\end{equation}
which can be written explicitly using~\eqref{eq: DiffLN representation}
\begin{equation}\label{eq: explicit ECS3 representation}
    \qty[U_{M,m}{(f,h,\alpha)}\Psi](p,\nu)=e^{i\oint \pair{p}{\alpha}+im \beta_W(f;\nu)+i\oint \pair{\nu}{x_W(h;p)}}\,\Psi((f,h)^{-1}\lga p,f^{-1}\lga \nu)\,,
\end{equation}
with the implicit definition $\Psi(p,\nu):=(\Psi(p))(\nu)$.

\subsection{Algebra action}

Let us now work out the action of the algebra generators on the wavefunction $\Psi(p,\nu)$, using the explicit form of the representation given in Eq.~\eqref{eq: explicit ECS3 representation}.

The action of translations (smeared with a function $\alpha\in L\RR^2$) is the simplest,
\begin{equation}
    \hat{P}[\alpha] \Psi(p,\nu) = \qty(\oint \pair{p}{\alpha})\Psi(p,\nu)\,,
\end{equation}
which can be written in polar coordinates to get a local version of~\eqref{eq: 2d polar coordinates}
\begin{equation}
    \hat P_0(\x) = e^{ s(\x)}\cos\Theta(\x), \quad \hat{P}_{1}(\x) = e^{ s(\x)}\sin\Theta(\x).
\end{equation}
In a similar manner to Lorentz transformations in the Poincaré representations, the generators of diffeomorphisms and $L\spl{2}$ have an external (or orbital) action and an internal one. The former comes 
from the transformation
of the momentum argument, $(f,h)^{-1}\lga p$, while the latter comes from the Wigner elements and the transformation of the $\Diff^{+}\!(S)/S$ coordinate, $f^{-1}\lga \nu$. Starting with $\Diff^{+}\!(S)$, we introduce a one-parameter family of diffeomorphisms connected to the identity
\begin{equation}
    f_t(\x) = \x + t\,\xi(\x) + \order{t^2}, \qquad f_t(\x)^{-1} = \x - t\,\xi(\x) + \order{t^2}.
\end{equation}
Using equation~\eqref{eq: momentum density transformation law}, we can write
\begin{equation}
    (f_t,e)^{-1}\lga p=(f_t^{-1})_* p = p + t(\xi p)' + \order{t^2},
\end{equation}
which shows that the first argument of the wavefunction is shifted by $t(\xi p)'$.
The orbital part of the generator can thus be written as
\begin{equation}
    \hat{L}_{\mathrm{orb}}[\xi] = - i \oint \left\langle(\xi p)',\fdv{}{p}\right\rangle =- i \oint \qty[\qty(\xi s' + \xi')(\x)\fdv{s(\x)} + (\xi \Theta')(\x)\fdv{\Theta(\x)}]\dd\x\,.
\end{equation}
The internal part is instead obtained by differentiating the phase $m \beta_W(f_t;\nu)$
\begin{equation}
    \beta_W(f_t;\nu)=\frac{2\pi}{M}\int^{t\xi(0)}_0 \nu(\x)^{1/3}\dd \x=\frac{2\pi}{M}\,t\,\xi(0)\nu^{1/3}(0)+\order{t^2},
\end{equation}
and the argument shift $f_{t}^{-1}\circ \nu$
\begin{equation}
    f_{t}^{-1}\circ \nu=\nu+t(\nu'\xi+3\nu\xi')+\order{t^2},
\end{equation}
so that
\begin{equation}
    \hat{L}_{\mathrm{int}}[\xi]\Psi(p,\nu)=\qty[-i\oint \left\langle \nu'\xi+3\nu\xi',\fdv{}{\nu}\right\rangle+\frac{2\pi}{M}\,m\,\xi(0)\nu^{1/3}(0)]\Psi(p,\nu)\,.
\end{equation}

We obtain the action of the $L\spl{2}$ generators in a similar way. For the orbital part, we write $h_t(\x) = \exp(t X(\x))$ and compute
\begin{equation}
    (\text{id},h_t)^{-1} \lga p= \sigma^*_{h_t^{-1}} p= p - t\,\Sigma_X^* p + \order{t^2},
\end{equation}
where $\Sigma^*_X$ is the matrix representation of the Lie algebra element $X$, the infinitesimal version of $\sigma^*_h$. This is just the per-point version of the $\spl{2}$ orbital action in the 2D case~\eqref{eq: 2d SL2R generators}, hence
\begin{subequations}
    \begin{align}
    \hat H_{\mathrm{orb}}(\x) &= i \cos{2 \Theta(\x)}\pqty{\fdv{}{s(\x)} -1} - i\sin{2\Theta(\x)}\fdv{}{\Theta(\x)}\,,\\
    \hat E_{\mathrm{orb}}(\x)  &= \frac{i}{2}\sin{2\Theta(\x)}\pqty{\fdv{}{s(\x)} -1} - i \sin^2{\Theta(\x)}\fdv{}{\Theta(\x)}\,,\\
    \hat F_{\mathrm{orb}}(\x) &= \frac{i}{2} \sin{2\Theta(\x)}\pqty{\fdv{}{s(\x)} -1} + i \cos^2{\Theta(\x)}\fdv{}{\Theta(\x)}\,.
\end{align}
\end{subequations}
For the internal part, we write the Wigner shearing as
\begin{equation}
    n_{x_W}=\sect{h}^{-1}_p\cdot e^{tX}\cdot \sect{h}_{\sigma_{e^{-tX}}^* p}=1+t(\Ad{\sect{h}^{-1}_p}X-\sect{h}^{-1}_p\cdot \pair{\Sigma^*_X p}{\nabla_p \sect{h}_{p}})+\order{t^2},
\end{equation}
so that $x_W(H)=2e^{-2s}\sin{2\Theta}$, and $x_W(E)=x_W(F)=-e^{-2s}\cos{2\Theta}$. This implies
\begin{subequations}
    \begin{align}
    \hat H_{\mathrm{int}}(\x)\Psi(p,\nu) &= 2\nu(\x) e^{-2s(\x)}\sin{2\Theta(\x)}\,\Psi(p,\nu)\,,\\
    \hat E_{\mathrm{int}}(\x)\Psi(p,\nu)  &= -\nu(\x) e^{-2s(\x)}\cos{2\Theta(\x)}\,\Psi(p,\nu)\,,\\
    \hat F_{\mathrm{int}}(\x)\Psi(p,\nu) &= -\nu(\x) e^{-2s(\x)}\cos{2\Theta(\x)}\,\Psi(p,\nu)\,.
\end{align}
\end{subequations}
These expression precisely reproduce the form of the internal component of the two-dimensional representation~\eqref{eq: 2d SL2R generators}. As the reader might have guessed --- perhaps guided by the title of the last section --- the \textit{Casimir} density $\nu\in\weight{3}$ is the eigenvalue of the $\ECS_2$ Casimir operator in these $\ECS_3$ representations
\begin{equation}
    \hat{\mc{C}}_{\mathrm{ECS}}(\x)\Psi(p,\nu) = \qty(-\hat{E}\hat{P_0}^2 + \hat{F}\hat{P}_1^2 + \hat{H}\hat{P_0}\hat{P_1})(\x)\Psi(p,\nu)= \nu(\x)  \Psi(p,\nu).
\end{equation}
On the other hand, only the integral of its cubic root is a Casimir of the full group
\begin{equation}
    \oint \hat{\mc{C}}^{1/3}_{\mathrm{ECS}}(\x)\Psi(p,\nu)\,\dd \x=M\Psi(p,\nu).
\end{equation}

\subsection{Unitarity}
\label{sec: measure}

In order to define a scalar product on the Hilbert space where the representation~\eqref{eq: implicit ECS3 representation} lives --- with the purpose of making the representation unitary --- we need two measures: the first one on the momentum orbit, the second one on the Casimir density orbit.
In finite-dimensional locally compact group, Mackey's theory ensures the existence of a quasi-invariant measure on the orbits, such that, after the proper Radon--Nikodym factor has been taken into account, the induced representation is invariant with respect to the $L^2$ scalar product induced by the measure \cite{Mackey1952} (see also \cite{Folland:2016} for a modern treatment). For infinite-dimensional groups like the one discussed in this paper, there are no general results on the existence of such a measure. However, if one can be constructed in some way, the unitarity of the induced representation still follows. In what follows, we will start by constructing a path-integral type measure on the momentum orbits. In what follows, we will start by constructing a path-integral-type measure on the momentum orbits. As usual in such a construction, all statements can be established rigorously for the regulated measure and are then conjectured to persist in the formal continuum limit. For the Casimir density orbit, we will make use of the work of Shavgulidze and Kosyak on quasi-invariant measures for diffeomorphism groups. Rather than attempting to define such a measure directly on the smooth orbit, we enlarge the configuration space to a lower-regularity completion, naturally described in terms of $\Diff^1_0(S)$, on which the Shavgulidze measure is well defined. The representation is then realized on the corresponding $L^2$ space, while the smooth diffeomorphism group continues to act as the physical symmetry group through its quasi-invariant action on this enlarged space. This provides a genuine measure-theoretic framework in which the Radon--Nikodym factors and the scalar product can be defined rigorously.

We start with the much simpler momentum orbit. Since the momentum density has two real components,
the natural candidate for a measure is the point-wise generalization of the Lebesgue measure used in the two-dimensional case. To give a precise meaning to this formal expression, we introduce a lattice regularization of the circle by choosing $N$ points
\begin{equation}
\label{eq: circle discretization}
    \x_i=\frac{2\pi i}{N}\,,
    \qquad
    i=0,\ldots,N-1,
\end{equation}
and define the finite-dimensional measure
\begin{equation}
    \dd\mu_N(p)
    \defeq
    \prod_{i=0}^{N-1}
    \dd p_0(\x_i)\,
    \dd p_1(\x_i).
\end{equation}
The functional measure on the momentum orbit is then formally understood as the continuum limit
\begin{equation}
    \mathcal Dp
    \defeq
    \lim_{N\to\infty}\dd\mu_N(p)
    =
    \prod_{\x\in S^1}
    \dd p_0(\x)\,\dd p_1(\x)
    \equiv
    \mathcal Dp_0\,\mathcal Dp_1.
\end{equation}
This is the standard formal flat functional measure for two real bosonic fields on the circle, of the same type as the integration measure over bosonic field configurations appearing in path-integral formulations of quantum field theory. At finite $N$, the action of $L_e\SL{2}$ is point-wise, so that the corresponding Jacobian is
\begin{equation}
    J_N[h]
    =
    \prod_{i=0}^{N-1}
    \left|\det \sigma_{h(\x)}^{*}\right|
    =1\,,
\end{equation}
where we used that $h(\x_i)\in\SL{2}$ for every lattice point. Therefore $\mathcal Dp$ is formally invariant under the loop group action in the continuum limit. To ensure diffeomorphism invariance, it is more convenient to work in Fourier space. We write
\begin{equation}
    p_a(\x) = \sum_{n\in \mathbb{Z}}p_{a,n} e^{in\x},\qquad \xi(\x) = \sum_{n\in\mathbb{Z}}\xi_n e^{in\x},
\end{equation}
with $p_{a,-n} = \bar{p}_{a,n}$ and $\xi_{-n} = \bar{\xi}_n$ to ensure reality. If we apply the infinitesimal diffeomorphism $f_t(\x) = \x + t\xi(\x)+\order{t^2}$ to $p$ and expand the transformed weight-one momentum, we find the Fourier modes
\begin{equation}
    (f_*p_a)_n = p_{a,n} -it\, n\sum_{m\in\mathbb{Z}}\xi_{n-m}p_{a,m}+\order{t^2}\,,
\end{equation}
which we can interpret as the application of an infinite-dimensional matrix $1+t M(\xi)$, with $[M(\xi)]_{n,m}= -in\xi_{n-m}$. Regularizing the measure by retaining only a finite number of Fourier modes $\abs{n}\leq N$ (unrelated to that in Eq.~\eqref{eq: circle discretization}), the Jacobian of the above transformation is
\begin{equation}
    J_N[f_t] = \det(1+ t M(\xi)) + \order{t^2} = 1 + t \tr[M(\xi)] + \order{t^2}.
\end{equation}
At finite $N$, it is easy to see that the trace of the matrix $M(\xi)$ vanishes. This result can be extended to all $\Diff^{+}\!(S)$ because of its path-connectedness. Given $f\in\Diff+(S)$, construct a path $t\mapsto f_t$ that starts at $\mathrm{id}$ and ends in $f$, and define the $t$-dependent vector field\footnote{The non-exponentiability of $\Diff^{+}\!(S)$, that is the lack of surjectivity of the exponential (even in a neighborhood of the identity), does not jeopardize the argument, as it only implies that there is no single $t$-independent vector field which is always tangent to the path $t\to f_t$.}
\begin{equation}
    \xi_t \defeq \dot f_t\circ f_t^{-1}\,.
\end{equation}
It is then easy to see that
\begin{equation}
    \frac{\dd}{\dd t}\log J_N[f_t]=\tr M(\xi_t)=0\,,
\end{equation}
where the last equality comes from the infinitesimal computation above. Since
$J[\mathrm{id}]=1$, it follows that $J[f]=1$ for any $f$, and therefore 
\begin{equation}
    \mathcal D(f\lga p)=\mathcal Dp\,.
\end{equation}
Setting aside the usual measure-theoretic subtleties associated with functional measures of the type appearing in path integrals, $\mathcal Dp$ is thus invariant under the full $\Diff^{+}\!(S)\ltimes L_e\SL{2}$ group action.
The standard induced-representation argument therefore applies and, provided an analogous invariant or quasi-invariant measure can be constructed on the Casimir-density orbit, the full representation will be unitary with respect to the corresponding $L^2$ inner product. 

We now move to the Casimir-density orbit. Contrary to the momentum orbit,
there is no natural candidate for an invariant flat functional measure. Additionally, there exists no quasi-invariant measure on non locally compact groups \cite{Weil:1965,Xia:1978}. 
Instead, we make use of the construction of Shavgulidze
\cite{Shavgulidze1978,Shavgulidze1988,Shavgulidze1995,Shavgulidze1997},
which provides Borel measures on finite-regularity diffeomorphism spaces that
are quasi-invariant under the left action of sufficiently smoother diffeomorphisms.\footnote{We denote by $\Diff^k_+\!(S)$ the group of orientation-preserving diffeomorphisms that are $C^k$.}
In particular, following the realization used by Kosyak \cite{Kosyak:1994}, we
consider the enlarged configuration space
\begin{equation}\label{eq: larger Casimir density orbit}
\widetilde{\mathcal O}_{\nu_\star}
    \cong
    \Diff^1_+\!(S)/\U{1}
    =
    \Diff^1_{+,0}(S),
\end{equation}
where the second is an equality between manifolds. We can equip these enlarged orbit with the Shavgulidze measure $\mu$ which is quasi-invariant under the
left action of $\Diff^2_+\!(S)$, that is
\begin{equation}\label{eq: Shavgulidze measure invariance}
    \mu(\mathcal A)=0
    \quad\Longleftrightarrow\quad
    \mu(f\rhd\mathcal A)=0
\end{equation}
for every $f\in\Diff^2_+\!(S)$ and every
$\mathcal A\in\mathcal B(\widetilde{\mathcal O}_{\nu_\star})$.\footnote{The
Borel $\sigma$-algebra $\mathcal B(X)$ is the smallest $\sigma$-algebra,
namely a collection of subsets closed under complements and countable unions
and intersections, containing all open sets of $X$.} Replacing $\Diff^\infty_+(S)$ with $\Diff^1_+\!(S)$ is not intended to enlarge the physical
symmetry group, but rather to provide a lower-regularity completion of the smooth Casimir-density orbit on which a genuine measure can be defined.
In particular,
the property~\eqref{eq: Shavgulidze measure invariance} holds for the smooth subgroup
$\Diff^\infty_+(S)\subset\Diff^2_+\!(S)$.
The Radon--Nikodym theorem then ensures that there exists a positive
measurable function $\rho_f(\nu)$ such that
\begin{equation}
    \dd(f_*\mu)(\nu)
    =
    \rho_f(\nu)\,\dd\mu(\nu).
\end{equation}
This function is called the Radon--Nikodym factor, or derivative. Using this result,
we can finally write the unitary representation of the $\ECS_3$ group by appending
the Radon--Nikodym factor to the induced representation
\eqref{eq: implicit ECS3 representation},
\begin{equation}\label{eq:fullrepresentation}
    \qty[
        U^{\mathrm{RN}}_{M,m}(f,h,\alpha)\Psi
    ](p,\nu)
    =
    \rho_f(\nu)^{\frac12}
    \qty[
        U_{M,m}(f,h,\alpha)\Psi
    ](p,\nu),
\end{equation}
where now $\nu\in\widetilde{\mathcal O}_{\nu_\star}$.
By construction, this representation is unitary with respect to the scalar
product
\begin{equation}
\label{eq: Hilbert space inner product}
    \braket{\Phi}{\Psi}
    =
    \int_{\mathcal O_{p_\star}}\mathcal Dp
    \int_{\widetilde{\mathcal O}_{\nu_\star}}
    \dd\mu(\nu)\,
    \overline{\Phi(p,\nu)}\,\Psi(p,\nu),
\end{equation}
where $\mathcal Dp$ is the invariant functional measure on the momentum
orbit constructed above and $\mu$ is the Shavgulidze measure on the enlarged
Casimir-density orbit. We denote by $\mathcal H$ the Hilbert space of $L^2$ wavefunctions with respect to this inner product. Notice that the scalar product is intrinsic to the
momentum orbit and the enlarged Casimir-density orbit and does not depend on
the choice of sections used to construct the induced representation. In
fact, a change of section modifies the wavefunctions by a point-dependent
little-group transformation which, being represented unitarily, leaves the
scalar product invariant.

\subsection{Irreducibility}
\label{sec: irreduibility}

We now address the issue of irreducibility of the representations constructed above. For second-countable locally compact groups, standard Mackey theory provides the appropriate framework for induced representations. In particular, for semidirect products with an abelian normal subgroup, the Wigner--Mackey little-group construction implies, under the usual regularity assumptions, that a representation induced from an irreducible little-group representation is itself irreducible. In the infinite-dimensional case we considered, the assumption of local compactness fails. Moreover, once we enlarge the Casimir-density orbit to~\eqref{eq: larger Casimir density orbit}, the action of the smooth symmetry group is no longer transitive, and irreducibility requires in particular that the action of the symmetry is ergodic.\footnote{A nonsingular action of a group $G$ on a measure space $(X,\mu)$ is \textit{ergodic} if every measurable subset $A\subset X$ that is invariant  (up to a measure zero subset) under the group action  has either measure zero or has a complement of measure zero.  In simpler words, there is no measurable subset of ``intermediate'' measure that is preserved by the group action.} Indeed, without this assumption, there would exist a nontrivial measurable subset of the orbit that is preserved by the group and wavefunctions supported on it would form a proper invariant subspace, thereby spoiling irreducibility.

In this section, we first establish an analogue of the Wigner--Mackey
irreducibility criterion for the representation $U_{M,m}$ of $\ECS_3$
in~\eqref{eq: implicit ECS3 representation}, and hence also for its unitary
realization~\eqref{eq:fullrepresentation}. More precisely, assuming that the
ergodicity of the regulated momentum orbit persists in the formal continuum
limit, we show that the full induced representation is irreducible whenever
the corresponding little-group representation $\overcirc U_{M,m}$ is
irreducible. We then show that the use of the enlarged Casimir-density orbit
allows us to establish the irreducibility of the little-group representation,
provided that the $\Diff^\infty_{+,0}(S)$ action on the enlarged
Casimir-density Hilbert space admits a strongly continuous extension to
$\Diff^2_{+,0}(S)$ with respect to the $C^2$ topology. Technical details are
deferred to Appendix~\ref{app:irreducibility}.

Our strategy is to show that any \textit{bounded} operator $T\in B(\mathcal H)$ on the representation Hilbert space $\mathcal H$ that commutes with the representation is necessarily a multiple of the identity. Schur's lemma then implies that the representation is irreducible. First, since $T$ commutes with translations, it has to preserve the momentum sector of the wavefunction,\footnote{In the infinite-dimensional setting, one has to put in additional work to prove the validity of this statement. See Appendix~\ref{app:C1} for the details.} namely
\begin{equation}
\label{eq: T acts as multiplication}
    [T\Psi](p;\nu) = T_p\Psi(p;\nu), \quad \mathrm{a.e.}\,
\end{equation}
where $T_p$ only acts on the remaining $\nu$ dependence and where equality almost everywhere (a.e.) means that it might fail on a subset of zero momentum measure $\mathcal{D}p$. The fact that $T$ also commutes with the representation of
$\Diff^{+}\!(S)\ltimes L_e\SL{2}$ implies that $T_p$ belongs to the commutant of the little-group representation, as shown in Appendix~\ref{app:C2}. Therefore, if the little group representation is irreducible --- as we take by assumption here --- we obtain
\begin{equation}\label{eq:Tisconstant}
    T_p \Psi(p;\nu)=c(p)\Psi(p;\nu) \quad \mathrm{a.e.},
\end{equation}
for some for some $c\in L^\infty(\mathcal O_{p_\star},\mathcal Dp)$.

In order to apply Schur's lemma to the full induced representation $U_{M,m}$, we need to show that $c(p)$ does not depend on $p$. As we show in Appendix~\ref{app:C3} using a path-integral argument, the action of $L_e \mathrm{SL}(2,\RR)$ on the momentum orbit $\mathcal O_{p_\star}$ is ergodic. Furthermore, if the action~\eqref{eq:Tisconstant} is to commute with the representation of $\Diff^{+}\!(S)\ltimes L_e\SL{2}$, as we assumed, $c(p)$ must be an invariant function
\begin{equation}
    c(p) = c(g^{-1}\rhd p) \quad \mathrm{a.e.}, \quad \forall g\in \Diff^{+}\!(S)\ltimes L_e\mathrm{SL}(2,\RR).
\end{equation}
As we show in Appendix~\ref{app:C4}, functions that are invariant under an ergodic action are constant almost everywhere, hence we deduce that $c(p)$ is a.e. independent of $p$. This concludes the proof of our first statement.

We now turn to the irreducibility of the little-group representation~\eqref{eq: DiffLN representation}. After the orbit enlargement, the representation is realized on
\begin{equation}
    \tilde{\mathcal K}=L^2(\tilde{\mathcal O}_{\nu_\star},\mu),
    \qquad
    \widetilde{\mathcal O}_{\nu_\star}
    \cong
    \Diff^1_+\!(S)/\U{1}=
    \Diff^1_{+,0}(S),
\end{equation}
where $\mu$ is the Shavgulidze measure in~\eqref{eq: Shavgulidze measure invariance}. As discussed above, this measure is
quasi-invariant under the action of $\Diff^2_0(S)$, so that the
diffeomorphism part of the representation extends to this larger group.
Let $A\in B(\mathcal K)$ be a bounded operator commuting with all operators of the original smooth little-group representation.
As before, the fact that it commutes with the normal subgroup $LN$ implies
\begin{equation}
    [A\psi](\nu)=a_\nu\psi(\nu)\,,
\end{equation}
with $a:\nu\mapsto a_\nu$ a measurable function on the enlarged orbit. Once again, the fact that $A$ is a multiplication needs to be proven, which is done in Appendix~\ref{app:C5}. 
Commutativity with the representations of based smooth diffeomorphisms implies
\begin{equation}
\label{eq: smooth A commutativity}
    a(\nu)=a(f^{-1}\rhd\nu)\quad \mathrm{a.e.},\quad \forall f\in\Diff^\infty_{+,0}(S). 
\end{equation}
 We now consider the action of $\Diff^2_0(S)$ on $\tilde{\mathcal K}$. Since $\Diff^\infty_{+,0}(S)$ is dense in $\Diff^2_{+,0}(S)$ in the $C^2$ topology, we can use the assumed strong continuity of the unitary representation to extend it to a representation of the larger group. Likewise, we can extend~\eqref{eq: smooth A commutativity} to\footnote{Indeed, if
$f_n\in\Diff^\infty_0(S)$ converges to $f\in\Diff^2_0(S)$ in the $C^2$
topology, then
\begin{equation}
    D_{f_n}\psi\longrightarrow D_f\psi
\end{equation}
in $\tilde{\mathcal{K}}$, and hence
\begin{equation}
    SD_f\psi
    =
    \lim_{n\rightarrow\infty}SD_{f_n}\psi
    =
    \lim_{n\rightarrow\infty}D_{f_n}S\psi
    =
    D_fS\psi.
\end{equation}
Consequently,
\begin{equation}
    a(\nu)=a(f^{-1}\rhd\nu)\quad \mathrm{a.e.},\quad \forall f\in\Diff^2_{+,0}(S). 
\end{equation}}
\begin{equation}
    a(\nu)=a(f^{-1}\rhd\nu)\quad \mathrm{a.e.},\quad \forall f\in\Diff^2_{+,0}(S). 
\end{equation}
Thanks to a result by Kosyak~\cite{Kosyak:1994}, according to which the Shavgulidze measure on $\Diff^1_{+,0}(S)$ is ergodic under the left action of $\Diff^2_{+,0}(S)$, we can deduce that $a$ is constant almost everywhere on $\tilde{\mathcal O}_{\nu_\star}$, and consequently that $A$ is a constant multiple of the identity on $\tilde{\mathcal K}$.
The original smooth little-group representation is therefore irreducible by Schur's lemma.

The two statements we proved in this section establish the irreducibility of the representation~\eqref{eq:fullrepresentation}.

\section{Maximal Central Extension of $\mathrm{ECS}_3$}
\label{sec: corner symmetry central extension}

Classically, the extended corner symmetry group $\ECS_3$ is expected to act faithfully on some corner phase space, and any central extension of it has no physical effect. Quantum mechanically, this is no longer true: projective representations are required by consistency of a the quantum description. This forces us to work with standard representations of the maximally centrally extended version of the symmetry group. Since $\ECS_3$ has both a $\Diff^{+}\!(S)$ factor and a $L\ECS_2$ component, one must in principle allow for independent central extensions, as well as central extensions of the semidirect product as a whole. The goal of this section is to determine such a maximal central extension.

As previously mentioned, the two-dimensional $\ecs_2$ algebra admits a central extension due to the non-trivial second group cohomology of $\RR^2$, namely $[P^0,P^1]=c$. Similarly, one can centrally extend $L\RR^2\to L\HH_3$ by tensoring the cohomology element with the symmetric pairing $(z^m,z^n)\mapsto \delta_{m+n,0}$. Explicitly:
\begin{equation}\label{eq: loop heisenberg}
    [P^0_m,P^1_n]= \delta_{m+n,0}\,c, \qquad c\;\,\text{central}.
\end{equation}
However, it is easy to show that this extension is incompatible with the action of $\diff(S)$. Starting from the Jacobi identity for the triple $(\ell_k,P_m^0,P_n^1)$
\begin{equation}
\label{eq: obstruction}
    [\ell_k,[P_m^0,P_n^1]]+[P_n^1,[\ell_k,P_m^0]]+[P_m^0,[P_n^1,\ell_k]]=0\,,
\end{equation}
we see that the first term vanish by the assumption that $c$ is central in $\ecs_3$, while the other two yield
\begin{equation}\label{eq: conformal obstruction to heisenberg}
    (m+n)\delta_{k+m+n,0}\,c=0\,,
\end{equation}
which is false for any $k\ne 0$ unless $c$ vanishes itself.

\paragraph{Conformal weights} The obstruction~\eqref{eq: conformal obstruction to heisenberg} has a clear interpretation in terms of conformal weights (the name come from the fact that the Witt algebra is the algebra of $2D$ conformal transformations). Assume that the $\SL{2}$ transformations and the $\RR^2$ transformations had two \textit{a priori} unrelated conformal weights $h$ and $h'$, that is
\begin{equation}
   [\ell_m, J^\alpha_n]=-(n+h m) J^\alpha_{m+n},\qquad [\ell_m, P_n^a]=-(n+h' m) P_{m+n}^a\,.
\end{equation}
Consistency of the algebra implies that $h=0$.\footnote{To check this, evalute the Jacobi identity for $(\ell_k,J^+_m,P_n^0)$ or any such triple with non-trivial structure constant between $J$ and $P$:
\begin{equation}
    [\ell_k,[J^+_m,P^0_n]]+[P^0_n,[\ell_k,J^+_m]]+[J^+_m,[P^0_n,\ell_k]]=0\implies \,
    h k P_{k+m+n}^1=0\nonumber
\end{equation}} Evaluating the Jacobi identity of Eq.~\eqref{eq: obstruction} now gives $(m+n+2h' k)$ rather than $(m+n)$ in~\eqref{eq: conformal obstruction to heisenberg}, which can be made into an identity assuming the conformal weight of translations is $h'=1/2$. A $\diff(S)\loplus (L\spl{2}\loplus L\RR^2)$ algebra with $(h,h')=(0,1/2)$ does indeed exist and admits a loop Heisenberg extension. The only reason we do not consider it is because of the physical derivation of the algebra $\ecs_3$, where normal special linear transformations (generated by $J_n^\alpha$) and normal translations (generated by $P_n^a$) transform in the same way under diffeomorphisms of the corner.

Let us now look for other central extensions by making the following ansatz on what the centrally extended bracket should look like
\begin{subequations}
\begin{align}
    &[\ell_m,\ell_n]=(m-n)\ell_{m+n}+E_{m,n}\,,\\
    &[\ell_m,V^A_n]=-n V^A_{m+n}+F^A_{m,n}\,,\\
    &[V^A_m,V^B_n]=C^{AB}{}_C V^C_{m+n}+D_{m,n}^{AB}\,,
\end{align}
\end{subequations}
where $E$, $F$ and $D$ are assumed to be commute with everything else and to have the expected symmetry properties. Since we are interested in non-trivial cocycles, we discard solutions which can be written as
\begin{equation}
    \phi_{m,n}(X,Y)=(m+n)\phi(X,Y)\,.
\end{equation}
We now use Jacobi identities to restrict the possible values of $E$, $F$ and $D$.
\begin{itemize}
    \item The Jacobi identity for three $l$ generators reveals that the only non-trivial $E_{m,n}$ is the Gelfand--Fuks cocycle $\coc_\V$, which leads to the well-known Virasoro extension:
    \begin{equation}\label{eq: Virasoro algebra cocycle}
    E_{m,n}=i\omega (\coc_\V)_{m,n}=\frac{\omega}{12}m^3\delta_{m+n,0}.
    \end{equation}
    Since the remaining generators form an ideal, this extension is automatically compatible with the others Jacobi identities.

    \item When considering three $V$ generators there are in principle four cases. Let us analyze them one by one: $(P,P,P)$ gives no condition as $\RR^2$ is abelian; $(P,P,J)$ yields a non-trivial constraint whose only solution is~\eqref{eq: loop heisenberg}, but we already excluded that in view of its incompatibility with $\diff(S)$; the condition on $(P,J,J)$ is only solved by trivial cocyles; $(J,J,J)$ yields the non-trivial condition
    \begin{equation}
        C^{\gamma\delta}{}_A D_{k,m+n}^{\beta A}+C^{\delta\beta}{}_A D_{m,n+k}^{\gamma A}+C^{\beta\gamma}{}_A D_{n,l+m}^{\delta A}=0.
    \end{equation}
    Since $C^{\beta\gamma}{}_a=0$, the sum only runs within $\spl{2}$. Given that $\spl{2}$ is a simple Lie algebra, the only solution to the above equation is the Kac--Moody extension:
    \begin{equation}\label{eq: Kac--Moody central extension}
        D^{\alpha\beta}_{m,n}=i \underline{k} (\coc_{\mathrm{KM}})_{m,n}^{\alpha\beta}=\underline{k}\,\kappa^{\alpha\beta}\, m \delta_{m+n,0}\,,
    \end{equation}
    with $\kappa$ the Killing form on $\spl{2}$. Given that the $\spl{2}$ generators have zero conformal weight, this extension is compatible with the action of $\diff(S)$. We underlined the central charge $\underline{k}$ to distinguish it from a Fourier index.

    \item In light of the previous findings, the mixed Jacobi identities imply that $F^A_{m,n}=0$.
\end{itemize}

Therefore, the maximal central extension of the $\ecs_3$ algebra is
\begin{equation}
    \qcs_3:=\mf{Vir}\loplus(\widehat{\spl{2}}\loplus L\RR^2)
\end{equation}
where $\mf{Vir}$ is the Virasoro algera and $\widehat{\spl{2}}$ is the affine extension of $L\spl{2}$. The group associated with this algebra is
\begin{equation}\label{eq: QCS3 definition}
    \QCS_3:=\Vir\ltimes (\widehat{\SL{2}}\times L\RR^2)\,,
\end{equation}
where $\Vir:=\widehat{\Diff}^{+}\!(S)$ is the Virasoro--Bott group and $\widehat{\SL{2}}$ is the maximal central extension of $\widetilde{L_e\SL{2}}$. We dub this the 3D \textit{quantum corner symmetry} group. In the following, we will apply Mackey's theory of induced representations to find (some of) its unitary irreducible representations.

\subsection{Group cocycles}

It is known that the Gelfand--Fuks cocycle $\coc_\V(\ell_m,\ell_n)=(\coc_\V)_{m,n}$ in~\eqref{eq: Virasoro algebra cocycle} integrates to the Bott--Thurston cocycle
\begin{equation}
    \COC_\V(f_1,f_2):=-\frac{1}{48\pi}\int_D \dd \log[(f_1^{-1})'] \wedge \dd \log[((f_1\circ f_2)^{-1})']\,,
\end{equation}
in the sense that $\delta\COC_\V=-\coc_\V$, where $\delta$ is the differential homomorphism between group cocycles and algebra cocycles.\footnote{Explicitly
\begin{equation}
\label{eq: Bott--Thurston cocycle}
    \delta \COC(X,Y):=\pdv[2]{}{t}{s} \bqty{\COC(e^{tX},e^{sY})-\COC(e^{sY},e^{tX})}\eval_{t=0,\,s=0}\,.
\end{equation}
} Notice that the integral is over a disk $D$ such that $\partial D=S$. The precise extension of the functions in the disk interior is not relevant, as the integral is always computed using Stokes' theorem: this is the reason why in~\eqref{eq: Bott--Thurston cocycle} we wrote derivatives with respect to $\x$ as if it was the only argument. Similarly, it is known that $\coc_\mathrm{KM}(J^\alpha_m,\,J^\beta_n)=(\coc_{\mathrm{KM}})_{m,n}^{\alpha\beta}$ integrates to the Mickelsson's cocycle~\footnote{One uses $\kappa^{\alpha\beta}=2\Tr(J^\alpha J^\beta)$.}
\begin{equation}
    \COC_{\mathrm{KM}}(h_1,h_2):=\frac{1}{8\pi}\int_D \Tr(h_1^{-1}\dd h_1 \wedge \dd h_2 h_2^{-1})\,.
\end{equation}
On the other hand, whereas the Bott--Thurston cocycle is a cocycle of the full group
\begin{equation}
    \COC_\V[(f_1,h_1,\alpha_1),(f_2,h_2,\alpha_2)]\equiv \COC_\V(f_1,f_2)\,,
\end{equation}
because the action of $\widetilde{\Diff}^{+}\!(S)$ on the universal cover of $L_e(\SL{2}\ltimes \RR^2)$ is semidirect, the Mickelsson's cocycle needs to be modified to make it compatible with the full group composition law:
\begin{equation}\label{eq: modified Mickelsson cocycle}
    \COC^{\Diff(S)}_\mathrm{KM}[(f_1,h_1,\alpha_1),(f_2,h_2,\alpha_2)]\equiv \COC_\mathrm{KM}(h_1,f_{1*}h_2)\,.
\end{equation}
No mixed cocycles appear because the first cohomology group of the universal cover of $L_e(\SL{2}\ltimes \RR^2)$ is trivial.
We can therefore claim that $\QCS_3$ in~\eqref{eq: QCS3 definition} is the maximal central extension of the group $\widetilde \ECS^+_3$ with $\RR$-valued extensions
\begin{equation}
    \QCS_3=\widetilde \ECS^+_3\times_{\COC_\V,\COC^{\Diff(S)}_{\mathrm{KM}}} (\RR_{\V}\times \RR_{\mathrm{KM}})\,.
\end{equation}

We now want to consider how these central extension modify the representation theory. As reviewed in Appendix~\ref{app: central extension theory}, a central extension modifies the group coadjoint action according to its Souriaru cocyle (cfr.~Eq.~\eqref{eq: coadjoint action centrally extended}). This discussion, applied to $\QCS_3$, tells us that we need to find the Souriau cocycles of $\COC_{\V}$ and $\COC^{\Diff(S)}_{\mathrm{KM}}$ to understand how the central elements affect the coadjoint orbits, which are equivalent to the orbits in the unitary dual. The first Souriau cocycle is given by the Schwarzian derivative $S[f]$, valued in $\diff(S)^*$:
\begin{equation}
\label{eq: Virasoro Souriau cocycle}
    \sCOC_{\V}[(f,h,\alpha)]=\sCOC_{\V}(f)=\qty(S[f](\dd\x)^2,0,0),\quad\text{where}\;\; S[f]=\frac{f'''}{f'}-\frac{3}{2}\pqty{\frac{f''}{f'}}^2,
\end{equation}
while the second gives (see Appendix~\ref{app:eq: modified Mickelsson cocyclecomputation} for a detailed derivation)
\begin{equation}
\label{eq: Kac--Moody Souriau cocycle}
S_{\KM}[(f,h,\alpha)] = \qty(3\Tr[L[f^*h]\otimes L[f^*h]],6\Tr[L[f^*h],\mdot],0).
\end{equation}

\subsection{First step: extended momentum density orbit}
Since both Souriau cocycles of $\QCS_3$ take values only in the Virasoro and Kac--Moody sectors (cfr.~Eqs.~\eqref{eq: Virasoro Souriau cocycle} and~\eqref{eq: Kac--Moody Souriau cocycle}), the momentum density orbits are not modified by the central extension.
We can therefore start Mackey's procedure from the same momentum representative
\begin{equation}
    p_\star = (1,0)\dd\x.
\end{equation}
The little group computed in the previous section, seen as a subgroup of $\QCS_3$ still stabilizes this momentum. Additionally, the two central elements act trivially on $p_\star$ and thus stabilize it as well. We therefore have
\begin{equation}
    \hat H_{p_\star} = \qty{(f,n_xa_f,\zv{},\zkm{} )\mid f\in \widetilde{\Diff}^{+}\!(S),\, n_x\in LN,\, \zv{}\in\RR_\V,\,\zkm{}\in \RR_\KM}
\end{equation}
as the little group inside $\QCS_3$.
Using the general product law~\eqref{eq: central extension composition law} we can compute the group product of two little group elements
\begin{equation}
\begin{split}
    (f_1, n_{x_1} a_{f_1}, \zv{1},\zkm{1})\cdot(f_{2}, n_{x_2} a_{f_2}, \zv{2},\zkm{2}))&=(f_1\circ f_2, n_{x_{12}} a_{f_1\circ f_2},\\
    &\qquad\zv{1}+\zv{2}+\COC_{\V}(f_1,f_2),\\
    &\qquad\zkm{1}+\zkm{2}+\COC_{\KM}(n_{x_1}a_{f_1},f_{1*}(n_{x_2}a_{f_2})),
\end{split}
\end{equation}
where again
\begin{equation}
    x_{12}(\x)=x_1(\x)+x_2(f_1^{-1}(\x))\pqty{\pdv{f_1^{-1}(\x)}{\x}}^{-2}\,.
\end{equation}
This shows that the little group in the centrally extended group can be rewritten as a central extension of the little group of the non-extended case
\begin{equation}
    \hat H_{p_\star} = \qty(\widetilde{\Diff}^{+}\!(S)\ltimes LN)\times_{\COC_V,\COC^{\Diff(S)}_{\KM}\eval_{H_{p_\star}}}(\RR_\V\times \RR_\KM)\,,
\end{equation}
where the extension is determined by the restriction of the group cocycles to the little group $H_{p_\star}=\Diff^{+}\!(S)\ltimes LN$ itself. We now evaluate this restriction. Denoting $h_1\equiv n_{x_1} a_{f_1}$ and $h_2\equiv n_{x_2} a_{f_2}$, we have
\begin{subequations}
    \begin{align}
    h_1^{-1}\dd h_1&=e^{2q_{f_1}} N\dd x_1-H \dd q_{f_1}\,,\\
    h_2^{-1}\dd h_2&=-H \dd q_{f_2}+N(2x_{2}\dd q_{f_2}+\dd x_2),
    \end{align}
\end{subequations}
where we introduced the shortcut notation $q_f:=\log(f^{-1})'$ and the $\spl{2}$ matrices
\begin{equation}
    N = \mqty(0&&1\\0&&0),\qquad H = \mqty(1&&0\\0&&-1).
\end{equation}
The cocycle dramatically simplifies thanks to $\Tr[N^2]=\Tr[HN]=0$ and $\Tr[H^2]=2$
\begin{equation}
    \COC_{\KM}(h_1,f_{1*}h_2)=\frac{1}{8\pi}\int_D\Tr[h_1^{-1}\dd h_1 \wedge f_{1*}(\dd h_2 h^{-1}_2)]= \frac{1}{4\pi}\int_D \dd q_{f_1} \wedge \dd (f_{1*}q_{f_2}).
\end{equation}
Note that the dependence on $x$ dropped from the expression. Therefore, this cocycle only modifies the group law of the $\Diff(S)$ sector. Using the relation $q_{{f_1}\circ {f_2}} = q_{f_1} + f_{1*}q_{f_2}$, we can write
\begin{equation}
  \COC_{\KM}^{\Diff(S)}\qty[(f_1,n_{x_1} a_{f_1}),(f_2,n_{x_2} a_{f_2})] = -12\, \COC_V(f_1,f_2).
\end{equation}
Therefore, upon performing the following change of central coordinates
\begin{equation}
    \overcirc z \defeq \zv{}\,, \qquad \overcirc w\defeq \zkm{} + 12\, \zv{}\,,
\end{equation}
the little group multiplication law can be stated as
\begin{equation}
    (f_1,x_1,\overcirc z_1,\overcirc w_1)\cdot (f_2,x_2,\overcirc z_2,\overcirc w_2)= \qty( f_1\circ f_2 ,x_{12}, \overcirc z_1 + \overcirc z_2 + \COC_V(f_1,f_2),\overcirc w_1+\overcirc w_2)
\end{equation}
which is that of the group 
\begin{equation}\label{eq: centrallyextendedlittlegroup}
    \hat H_{p_\star} = (\Vir\ltimes LN)\times \RR\,.
\end{equation}

\subsection{Second step: extended Casimir density orbit}
The little group~\eqref{eq: centrallyextendedlittlegroup} differs from the non-extended one of Section~\ref{sec: 3D case} by two factors of $\RR$. One of them is in a trivial Cartesian product, the other is the central term of the Bott--Virasoro group and thus modifies the little group computation. For simplicity, we drop the former and focus on
\begin{equation}
    \hat H^0_{p_\star} = \Vir\ltimes LN.
\end{equation}
Since the Bott--Thurston cocycle only enters in the group multiplication of elements of $\Diff^{+}\!(S)$, the representations of $LN$ and their orbit under the action of $\Diff^{+}\!(S)$ are the ones given in Section~\ref{sec: Casimir density orbit}. Note that for vanishing Casimir density $\nu=0$, the stabilizer is the entire Bott--Virasoro group, whose unitary irreducible  highest-weight representations are obtained from unitary irreducible quotients of Verma modules \cite{GoddardKentOlive:1986,GoodmanWallach:1985}. Let us work out the nowhere vanishing density case in details. As we already know, the orbits of the Virasoro dual action in $LN^* = \mathcal{E}[3]$ are characterized by the invariant
\begin{equation}
    M = \oint \nu^{1/3},\quad \nu \in \mathcal{E}[3].
\end{equation}
Therefore, as in the previous section, we can pick the constant representative
\begin{equation}
    \nu_\star = \frac{M^3}{8\pi^3}(\dd\x)^3.
\end{equation}
In the extended case, the little group of $\nu_\star$ also contains the
central Virasoro subgroup. Since the Bott--Thurston cocycle vanishes when
restricted to rigid rotations,
the lifted rotations do not mix with the central coordinate. The preimage
of the rotation subgroup in the Virasoro group is therefore the direct
product of the lifted rotation subgroup with the center
\begin{equation}
    \Vir_{\nu_\star}
    =
    \widetilde{\U{1}}\times\mathbb R_{\V}
    \cong \mathbb R^2,
\end{equation}
where we replaced the group of rigid rotations $\U{1}$ by its universal cover ($\cong\RR$) since we are working in $\widetilde\Diff^{+}\!(S)$. The Casimir density orbits are consequently unchanged by the central extension:
\begin{equation}
\Diff^{+}\!(S)/S\longrightarrow \Vir/H_{\nu_\star}
\cong
\widetilde{\Diff}^{+}(S)/\widetilde{\U{1}}.
\end{equation}
Unitary irreducible representations of $\Vir_{\nu_\star}$ are simply given by 
\begin{equation}
    \tau_{m,\omega_{\mathrm{eff}}}(r_\beta,\overcirc z) = e^{i(m\beta+\omega_{\mathrm{eff}} \overcirc z)},
\end{equation}
where $\omega_{\mathrm{eff}}\in \RR$ is the central charge of the Virasoro algebra inside the little group. From the above construction, we have
\begin{equation}
    \omega_{\mathrm{eff}} = \omega - 12 \underline{k}\,.
\end{equation}  

Following Mackey's procedure, we extend $\tau_{m,\omega_{\mathrm{eff}}}$ to a representation of
$\Vir_{\nu_\star}\ltimes LN$ by tensoring it with the $LN$ character associated with $\nu_\star$, and then induce this representation to the full little group $\hat H^0_{p_\star}=\Vir\ltimes LN$.
The induced representation acts on functions of the orbit
\begin{equation}
\psi:
\Vir/H_{\nu_\star}
\longrightarrow \CC.
\end{equation}
As explained in Section~\ref{sec: Casimir density orbit}, we can parametrize every Casimir density $\nu$ in the same orbit as $\nu_\star$ as $\sect{f}_\nu\lga \nu_\star$, where $\sect{f}_\nu$ is the based diffeomorphism defined in~\eqref{eq: standard based diffeo}. At the level of the Bott--Virasoro group, we choose the section with vanishing central coordinate,\footnote{\label{foot: vanishing central extension lift} Of course, since the central extensions acts trivially on the dual space, any lift would bring $\nu_\star$ to $\nu$. The vanishing central extension choice is simply convenient.} i.e.
\begin{equation}
    \widehat{\sect{f}}_\nu:=(\sect{f}_\nu,0)\in\Vir\,.
\end{equation}
The corresponding Virasoro Wigner element is thus
\begin{equation}
    \widehat W(f,\overcirc z;\nu):=\widehat{\sect{f}}_\nu^{-1}\cdot (f,\overcirc z)\cdot \widehat{\sect{f}}_{(f,\overcirc z)^{-1}\lga\nu}\in \Vir_{\nu_\star}\,.
\end{equation}
Its $\widetilde{\U{1}}$ component is $\sect{f}_\nu^{-1}\cdot f\cdot\sect{f}_{(f,\overcirc z)^{-1}\lga\nu}$ which is the same as $\sect{f}_\nu^{-1}\cdot f\cdot\sect{f}_{f^{-1}\lga\nu}$ given that the central element does not affect $\nu$. We recognize the same expression of the non-extended case~\eqref{eq: U(1) little group element}, namely $r_{\beta_W}$ with
\begin{equation}
    \beta_W(f;\nu)=\frac{2\pi}{M}\int_0^{f(0)}\nu^{1/3}(\x)\dd\x\,,
\end{equation}
now seen as a translation in $\widetilde{\U{1}}$. The central component of $\widehat{W}$ is
\begin{equation}
    \overcirc z_W(f,\overcirc z;\nu)=\overcirc z+\COC_{\V}(\sect{f}^{-1}_\nu,f)+\COC_{\V}(\sect{f}^{-1}_\nu\circ f, \sect{f}_{(f,\overcirc z)\lga\nu})\,,
\end{equation}
which can be written, for compactness, in terms of the section-dependent combination
\begin{equation}
\label{eq: Virasoro Ludovic cocycle}
    \Gamma_{\V}(f;\nu):=\COC_{\V}(\sect{f}^{-1}_\nu,f)+\COC_{\V}(\sect{f}^{-1}_\nu\circ f, \sect{f}_{(f,\overcirc z)\lga\nu})\,.
\end{equation}

The induced representation of $\Vir\ltimes LN$ can therefore be written
\begin{align}
\qty[\overcirc U_{M,m,\omega_{\mathrm{eff}}}{(f,x,\overcirc z)}
\psi](\nu)=
\exp\qty[i\omega_{\mathrm{eff}}\overcirc z_W+i m\beta_W+i\oint \pair{\nu}{x}
]
\psi((f,\overcirc z)^{-1}\lga \nu),
\end{align}
which generalizes the representation~\eqref{eq: DiffLN representation} and reduces to it in the case of vanishing Virasoro effective central charge $\omega_{\mathrm{eff}}=0$.

Next, we tensor this representation with the characters of the trivial $\RR$ factor in equation~\eqref{eq: centrallyextendedlittlegroup}
\begin{equation}
    \overcirc U_{M,m,\omega,\underline{k}}{(f,x,\overcirc z,\overcirc w)}\psi=e^{i\underline{k}\overcirc w}\overcirc U_{M,m,\omega_{\mathrm{eff}}}{(f,x,\overcirc z)}\psi\,.
\end{equation}
This operation essentially produces selection rules between sectors with different values of $\underline{k}$. Then we further tensor in the characters of the momentum orbits to form the representation that we will induce to the rest of $\QCS_3$. In order to do so we consider the section  
$(\text{id},\sect{h}_p) \in \widetilde{\Diff}^{+}\!(S)\ltimes L\spl{2}$  such that $p = (\text{id},\sect{h}_p)\lga p_\star$ given in~\eqref{eq: momentum orbit parametrization} and lift it to the full group $\Vir\ltimes \widehat{L\spl{2}}$ by choosing vanishing central extensions (cfr.~footnote~\ref{foot: vanishing central extension lift})
\begin{equation}
    \widehat{\sect{h}}_p \defeq (\mathrm{id},\sect{h}_p,0,0)\in \Vir\ltimes \widehat{L\spl{2}}\,.
\end{equation}
We now ask how the Wigner element~\eqref{eq: Wigner element for momentum orbits} is modified by the presence of central extensions. Because of the trivial action of the central elements on $p$, the $\widetilde{\ECS}^+_3$ component of the Wigner element is untouched, hence we recover $(f_W,n_{x_W}a_{f_W})$, with $f_W=f$ and $x_W\in\weight{-2}$ as in~\eqref{eq: Wigner element for momentum orbits}. Moreover, given that the section $\widehat{\sect{h}}_p$ has trivial diffeomorphism and central components, the Bott--Thurston cocycle does not contribute and we have
\begin{equation}
    z_W(g;p)=z\,,\qquad g\in\Vir\ltimes \widehat{L\spl{2}}.
\end{equation}
The only change appears in the Mickelsson cocycle. A straightforward computation gives
\begin{equation}
    {\zkm{}}_W(g;p) = \zkm{} + \COC_{\KM}(\sect{h}^{-1}_p,h) + \COC_{\KM}(\sect{h}_p^{-1}h,f_* \sect{h}_p')
    =: \zkm{} + \Gamma_{\KM}(g;p),
\end{equation}
where we defined the shortcut
\begin{equation}
    \Gamma_{\KM}(g;p) \defeq \COC_{\KM}(\sect{h}^{-1}_p,h) + \COC_{\KM}(\sect{h}_p^{-1}h,f_* \sect{h}_p')
\end{equation}
similarly to~\eqref{eq: Virasoro Ludovic cocycle}.
The final step of the induction eventually provides the $\mathrm{QCS}_3$ representation
\begin{equation}
\label{eq: QCS3 reps}
    \qty[U_{M,m,\omega,\underline{k}}{(g,\alpha)}\Psi](p,\nu)=e^{i\oint\pair{p}{\alpha}+i\underline{k}{\zkm{}}_W+i\omega{\zv{}}_W+im\beta_W+i\oint\pair{\nu}{x_W}}\Psi(g^{-1}\lga p,(f,z)^{-1}\lga \nu)
\end{equation}
where $g=(f,h,\zv{},\zkm{})$ is a $\Vir\ltimes \widehat{L\spl{2}}$ element. Note the following cancellation in the exponent
\begin{equation}
    \underline{k}\overcirc w+\omega_{\mathrm{eff}}\overcirc z=\underline{k}(\zkm{}+12\zv{})+(\omega-12\underline{k})\zv{}=\underline{k}\zkm{}+\omega\zv{}
\end{equation}

Determining the representation~\eqref{eq: QCS3 reps} is the main result of this work. The wavefunction $\Psi$ represents the quantum state of a certain corner configuration of gravity (in a suitably extended phase space), which is thus parametrized by two densities over the corner, one with weight $1$ and one with weight $3$. We believe that these wavefunctions, being in representations of the (quantum) corner symmetry group, are worth investigating more in the context of the corner proposal.

\paragraph{Unitarity and irreducibility}

The inclusion of the central extensions does not alter the general structure of the unitarity and irreducibility arguments. The additional central terms enter the representation only through phase factors and therefore do not affect the scalar product nor the Radon--Nikodym factor --- provided that these phases are well-defined and measurable on the enlarged Casimir-density space. Likewise, once an operator commuting with the little-group representation has been reduced to a multiplication operator, the phases commute with it and drop out of the irreducibility argument.
The only additional subtlety is therefore one of regularity: the Bott--Thurston cocycle~\eqref{eq: Bott--Thurston cocycle} involves derivatives that are not obviously defined for generic elements of the $C^1$ completion. The extension of the representation~\eqref{eq: QCS3 reps} to the enlarged orbit requires a separate regularity analysis, although no new obstruction is expected at the level of the commutant once such an extension is established.

\section{Conclusion}
\label{sec: end}
The present work should be regarded as a first step towards understanding the
representation theory of three-dimensional corner symmetries, rather than as
an exhaustive classification. Indeed, the usual completeness statements of
the Wigner--Mackey construction rely on hypotheses that are not immediately available for
the infinite-dimensional, non-locally compact groups considered here. In
addition, throughout this work we restricted ourselves to the regular duals
of the relevant loop spaces, describing their elements by smooth or continuous
densities. The full continuous dual also contains genuinely distributional
configurations, including finite sums of delta-function contributions, whose
orbits and little groups may lead to qualitatively different representations.
Such distributional sectors are particularly interesting from the perspective
of recent proposals in which localized ``embadons'' provide elementary
degrees of freedom for quantum geometry \cite{Ciambelli:2024swv}. Therefore, extending the present analysis beyond the regular dual and determining whether such localized configurations admit a natural interpretation within the three-dimensional corner framework constitute important directions for future work.

A first remarkable feature of the representation theory uncovered here is
that several structures that have appeared independently in the study of
three-dimensional quantum gravity arise naturally as particular sectors of
the corner-symmetry construction. For instance, the trivial momentum orbit
$\mathcal O_{p=0}$ admits the full $\Diff^+\!(S)\ltimes L\SL{2}$ symmetry as its little group and is thus naturally
related to the $\SL{2}$ WZW representations familiar from the
Chern--Simons formulation of three-dimensional gravity \cite{Witten:1988hc,CoussaertHenneauxVanDriel:1995}. For the non-vanishing momentum orbit studied in detail in this work, the
little group~\eqref{eq: 3D little group}
is instead quite similar to the three-dimensional BMS group, with the
main difference being the weight of its abelian normal subgroup.
In the centrally
extended theory, the vanishing Casimir-density orbit $\mathcal O_{\nu=0}$ has the
full Virasoro group as its little group, thereby bringing the usual unitary
representations of Virasoro directly into the corner-symmetry framework.
It is striking and definitely noteworthy that WZW, BMS-like, and Virasoro representation
theory, all of which play prominent roles in different approaches to
three-dimensional gravity, reappear here as different sectors or limiting
cases of the representation theory of the same local corner symmetry group.
This suggests that the corner-symmetry framework may provide a common setting in which these seemingly distinct
structures can be understood and compared.

A particularly interesting direction is to investigate the physical
information encoded in the representation-theoretic structures uncovered
here. Several of the little groups that arise in the construction possess
Virasoro- or, more generally, CFT-like sectors, suggesting that their
representation theory may provide a direct route to the counting of corner
degrees of freedom. In particular, whenever a Virasoro sector with non-zero
central charge is present, one may ask whether the asymptotic density of
states can be extracted through Cardy-type arguments and related to the
entropy associated with a gravitational corner. Similarly, the close
relation between the little group identified in
Eq.~\eqref{eq: 3D little group} and $\mathrm{BMS}_3$ suggests investigating
whether the BMS version of the Cardy formula~\cite{Bagchi:2012xr}, or a suitable generalization
adapted to the different weight of the abelian sector, can be used
to count states in the corresponding representations. The quantum numbers entering such counting formulas should in turn admit a geometric
interpretation through the corresponding classical moment maps: evaluating
the corner charges on interesting gravitational configurations, such as
the BTZ black hole, would relate the quantum representation data to
geometric quantities such as the horizon size and angular momentum. This would provide a concrete setting in which to test whether the
entropy obtained from the corner Hilbert space reproduces the gravitational
entropy, and, more ambitiously, whether it can be interpreted as the
entanglement entropy associated with the partition of spacetime across
the corner.

Lastly, the present construction may provide a natural setting in which to revisit the problem of Hilbert space factorization in gravity. As emphasized in the foundational work of Donnelly and Freidel~\cite{donnelly_local_2016}, gauge invariance obstructs a naive factorization of the Hilbert space across a spatial boundary, since the two regions share boundary degrees of freedom and are subject to matching gauge constraints at the corner. The extended phase space construction resolves this tension by promoting the would-be gauge transformations at the boundary to physical edge modes. In the two-dimensional case, this idea was developed in~\cite{ciambelli_quantum_2024,varrin_physical_2024,ciambelli_entanglement_2026-2,Varrin:2025okc,Kowalski-Glikman:2025arealaw}, where the representation theory of the quantum corner symmetry group was used to formulate the gluing condition of adjacent quantum regions and to characterize the corresponding entanglement structure. It would therefore be particularly interesting to understand whether the three-dimensional representations constructed here admit an analogous gluing prescription, and whether their additional functional degrees of freedom lead to new universal contributions to the entanglement entropy associated with the gravitational field.

\section*{Acknowledgment}
We thank Luca Ciambelli and Jerzy Kowalski-Glikman for their great help and mentorship throughout the project. We also thank Rodrigo Andrade e Silva, Simon Langenscheidt, Laurent Freidel, Glenn Barnich, and Jose M. Figueroa-O'Farrill for useful discussions.
GN is supported by the 1st edition (2025) of ``Research fellowship in fundamental physics and study of the universe'' established by the Antonio Madonna Foundation ETS. GN is also grateful for the hospitality of Perimeter Institute where part of this work was carried out.

\appendix

\section{Loop group theory}
\label{app: loop group theory}

We write an element of $\ECS_3^+$ as
\begin{equation}
    (f,h,\alpha),\quad f\in\Diff^{+}\!(S),\; h\in L_e\SL{2},\; \alpha\in L\RR^2\,.
\end{equation}
The composition law is
\begin{equation}
    (f_1,h_1,\alpha_1)\cdot (f_2,h_2,\alpha_2)=(f_1\circ f_2,\, h_1\cdot f_{1*}h_2,\, \alpha_1+\sigma_{h_1}f_{1*} \alpha_2),
\end{equation}
where $\sigma_h$ denotes the fundamental ($2\times 2$ matrix) representation of $\SL{2}$ on $\RR^2$. Since $f$ is a diffeomorphism, and thus invertible, the push-forward of a function by $f$ is equivalent to its pull-back by $f^{-1}$. 

We can easily check that the inverse of any element is
\begin{equation}
\label{eq: loop group inverse}
    (f,h,\alpha)^{-1}=(f^{-1},f^*h^{-1},-\sigma^{-1}_{f^*h}f^*\alpha).
\end{equation}
Therefore, using $(f_1\circ f_2)_*=(f_2^{-1}\circ f_1^{-1})^*=f_1^{-1*}f_2^{-1*}=f_{1*}f_{2*}$, conjugation gives
\begin{equation}
\begin{split}
    &(f_1,h_1,\alpha_1)\cdot (f_2,h_2,\alpha_2)\cdot(f_1,h_1,\alpha_1)^{-1}=\\
    &=(F,\,h_1\cdot f_{1*}h_2 \cdot F_* h_1^{-1},\,\alpha_1+\sigma_{h_1}f_{1*} \alpha_2-\sigma_{h_1}\sigma_{f_{1*}h_2} \sigma^{-1}_{F_*h_1}F_*\alpha_1), 
\end{split}
\end{equation}
where $F\equiv f_1\circ f_2 \circ f_1^{-1}$.
From this, writing the middle element as the exponential of a Lie algebra element, we obtain the adjoint action of $\ECS_3$ on $\ecs_3$:
\begin{equation}
    \Ad{(f,h,\alpha)}(\xi,J,b)=(f_* \xi,\,\Ad{h}f_*J+\iota_{f_*\xi}R[h],\,\sigma_h f_*b-\Sigma_{\Ad{h}f_*J+\iota_{f_*\xi} R[h]}\alpha+\iota_{f_*\xi}\dd\alpha).
\end{equation}
In the above, we introduced the right-invariant $1$-form $R[h]=\dd h\cdot h^{-1}$, valued in $L\spl{2}$.\footnote{The invariance is under right translation by constant $\SL{2}$ elements.}

The natural pairing $\ecs_3^*$ is defined by
\begin{equation}
    \pair{(\mu,y,p)}{(\xi,J,b)}=\oint(\iota_\xi \mu+\pair{y}{J}+\pair{p}{b})\,,
\end{equation}
where the pairings on the r.h.s.~are respectively that of $\SL{2}$, $\pair{y}{J}=\Tr(yJ)$, and that of $\RR^2$, $\pair{p}{b}=p\cdot b$. Dualizing with respect to this, we obtain the coadjoint action of $\ECS_3$
\begin{equation}
    \coAd{(f,h,\alpha)}(\mu,y,p)=(f_* \mu-\pair{f_* y}{L[h]}-\pair{\sigma^*_hf_*p}{\dd\alpha},\,\coAd{h}f_*y+\alpha\times \sigma^*_h f_* p,\,\sigma^*_{h} f_* p),
\end{equation}
where $\pair{\alpha\times p}{J}=\pair{p}{\Sigma_J\alpha}$ and we used $\coAd{h}(\alpha\times p)=\sigma_h\alpha\times \sigma^*_h p$.

The action of $\Diff(S)\ltimes L\SL{2}$ on the dual space $L\RR^{2*}$ is thus
\begin{equation}
\label{eq: coadjoint action of DiffLSL}
    \coAd{(f,h)}p=\sigma^*_{h} f_* p\,.
\end{equation}
As explained in the main text (see Eq.~\eqref{eq: 3D little group}), the stabilizer $H_{p_\star}$ of a constant momentum density $p_\star=(1,0)\dd\x$ is given by elements of the form
\begin{equation}
    (f,n_x a_f)
\end{equation}
where, for any $\x\in S$, $a_{f}(\x)$ and $n_{x}(\x)$ are elements of $\SL{2}$ whose $\sigma^*$ representation respectively coincide with an element of $A$ and $N$ in the Iwasawa decomposition~\eqref{eq: KAN decomposition}. In particular, $a_f(\x)$ is given by Eq.~\eqref{eq: stabilizing boost}. The proof takes one line
\begin{equation}
    \coAd{(f,n_x a_f)}p_\star=\sigma^*_{n_x a_f}(f_* p_\star)=\mqty(\bqty{(f^{-1})'(\x)}^{-1}&0\\0&(f^{-1})'(\x))\mqty(1\\0)\pdv{f^{-1}(\x)}{\x}\dd\x=\mqty(1\\0)\dd\x=p_\star\,.
\end{equation}
Stripping the $L\RR^2$ component from the $\ECS_3$ composition law, we determine the product of two elements in $H_{p_\star}$
\begin{equation}
    (f_1,n_{x_1}a_{f_1})\cdot (f_2,n_{x_2}a_{f_2})=(f_1\circ f_2,\, 
    n_{x_1}a_{f_1}\cdot f_{1*}n_{x_2}a_{f_2}).
\end{equation}
We can compute explicitly the second entry using the identity
\begin{equation}
    \mqty(1&x_1\\0&1)\mqty(\rho_1&0\\0&\rho^{-1}_1)\mqty(1&x_2\\0&1)\mqty(\rho_2&0\\0&\rho^{-1}_2)=\mqty(1&x_1+x_2\rho_1^2\\0&1)\mqty(\rho_1\rho_2&0\\0&\rho^{-1}_1\rho^{-1}_2)\,
\end{equation}
so that
\begin{equation}
\label{eq: little group composition law}
    n_{x_1}a_{f_1}\cdot f_{1*}n_{x_2}a_{f_2}=n_{x_{12}} a_{f_1\circ f_2}
\end{equation}
with
\begin{equation}
    x_{12}(\x)=x_1(\x)+x_2(f_1^{-1}(\x))\bqty{(f_1^{-1})'(\x)}^{-2}\,.
\end{equation}
This proves that $H_{p_\star}$ is indeed a subgroup.

\section{Central extensions}
\label{app: central extension theory}

Given a Lie group $G$, its central extension is a deformation of the trivial product $G\times \RR$, allowed by the existence of non-trivial two-cocycles on $G$. If we let $\COC\in H^2(G,\RR)$ be any such cocycle, the composition law for the centrally extended group $\hat G$ is
\begin{equation}
\label{eq: central extension composition law}
    (g_1,z_1)\cdot(g_2,z_2)=(g_1\cdot g_2,z_1+z_2+\COC(g_1,g_2)),
\end{equation}
where the cocycle property $\COC(g_1,g_2)+\COC(g_1\cdot g_2,g_3)=\COC(g_1,g_2\cdot g_3)+\COC(g_2,g_3)$ ensures that the group composition is associative.
Since the inverse is
\begin{equation}
    (g,z)^{-1}=(g^{-1},-z),
\end{equation}
given that, for a normalized cocycle, $\COC(g,g^{-1})=0$, we can easily work out the conjugation
\begin{equation}
    (g_1,z_1)\cdot(g_2,z_2)\cdot(g_1,z_1)^{-1}=(g_1\cdot g_2\cdot g_1^{-1},z_2+\COC(g_1,g_2)+\COC(g_1\cdot g_2,g_1^{-1})).
\end{equation}

Pushing-forward the conjugation, we can write the adjoint action of the centrally extended group $\hat G$ on $\hat\g=\g\oplus \RR$
\begin{equation}\label{eq: adjoint action centrally extended}
    \Ad{(g,z)}(X,\mf s)=\qty(\Ad{g}X,\mf s-\frac{1}{24\pi}\pair{\sCOC[g]}{X}),
\end{equation}
where $\sCOC: G\to \g^*$ is the Souriau cocycle associated with $\COC$, defined by
\begin{equation}\label{eq: sourieaudefinition}
    \dv{}{t}\qty(\COC(g,e^{tX})+\COC(g\cdot e^{tX},g^{-1}))\eval_{t=0}=:-\frac{1}{24\pi}\pair{\sCOC[g]}{X}\,.
\end{equation}
Taking another push-forward defines the algebra adjoint action, which coincide with the Lie bracket
\begin{equation}
    \ad{(X,\mf s)}(Y,\mf t)=[(X,\mf s),(Y, \mf t)]=([X,Y],\,-\delta \COC(X,Y)),
\end{equation}
where we introduced the differential $\delta\COC$ of the group cocycle, which defines a Lie algebra cocycle:
\begin{equation}
    \pdv[2]{}{t}{s}\qty(\COC(e^{tX},e^{sY})-\COC(e^{sY},e^{tX}))\eval_{t=0,s=0}=:\delta \COC(X,Y).
\end{equation}
In particular, the $\RR$ generator commutes with everything, thus the name `central extension'
\begin{equation}
    [(X,\mf s),(0,\mf t)]=0\,.
\end{equation}

The dual of the centrally extended Lie algebra is $\hat\g^*=\g^*\otimes \RR^*$, defined by the pairing
\begin{equation}
    \pair{(m,\mf u)}{(X,\mf s)}
    =\pair{m}{X}+\mf u\, \mf s,\quad \forall m\in\g^*,\,X\in \g,\,\mf u\in\RR^*,\,\mf s\in \RR\,.
\end{equation}
Dualizing the adjoint action, the coadjoint action of the centrally extended group is
\begin{equation}\label{eq: coadjoint action centrally extended}
\coAd{(g,z)}(m,\mf u)=\qty(\coAd{g}m-\frac{1}{24\pi}\mf u\, \sCOC[g^{-1}],\mf{u})\,.
\end{equation}
In particular, the central component of the coadjoint vector is invariant under the coadjoint action, $\mf u\mapsto \mf u$. From this, we can also deduce the algebra coadjoint action
\begin{equation}
\coad{(X,\mf s)}(m,\mf u)=\qty(\coad{X}m-\mf u\, \delta\COC[X,\mdot],\mf{u})\,.
\end{equation}

\subsection{Computation of the Kac--Moody Sourieau cocycle}
\label{app:eq: modified Mickelsson cocyclecomputation}

In order to compute the Souriau coycle associated with the modified Mickelsson's cocycle $\COC_{\KM}^{\Diff(S)}$, we differentiate equation~\eqref{eq: modified Mickelsson cocycle} according to equation~\eqref{eq: sourieaudefinition}. We denote $g=(f,h,\alpha)\in \mathrm{ECS}_3$ and $ X= (\xi,J,b)\in \mathfrak{ecs}_3$ and also write
\begin{equation}
    e^{tX}=(f_t,h_t,\alpha_t).
\end{equation}
The computation of the Souriau cocycle requires $\COC_{\KM}^{\Diff(S)}(g,e^{tX})$, which we can easily compute at first order
\begin{equation}
     \COC^{\Diff(S)}_\KM(g,e^{t X}) = \COC_{\KM}(h,f_{*}(1 + t J + \order{t^2})= \COC_{\KM}(h,1 + t f_*J + \order{t^2}).
\end{equation}
Writing $k_t \defeq 1 + t f_* J$, we can compute $\dd k_t k_t^{-1} = t \dd (f_* J)+\order{t^2}$
and thus 
\begin{equation}
\label{eq: Souriau computation first term}
\dv{t}\COC^{\Diff(S)}_{\KM}(g,e^{tX})\eval_{t=0} = \frac{1}{8\pi}\int_D\Tr[h^{-1}\dd h\wedge\dd(f_{*}J)].
\end{equation}
Next we need to compute $\COC^{\Diff{S}}_{\KM}(g\cdot e^{tX}, g^{-1})$. To do this, we need to extract the $\Diff^{+}\!(S)$ and $L_e\SL{2}$ components of $g\cdot e^{tX}$, which we respectively denote by $F_t$ and $H_t$. For the second argument, we also need the group inverse~\eqref{eq: loop group inverse}. We then have
\begin{equation}
    \COC^{\Diff(S)}_{\KM}(g\cdot e^{tX}, g^{-1})=\COC_{\KM}(H_t,F_{t*}(f^* h^{-1})).
\end{equation}
Again, at first order in $t$, we use the $\ECS_3$ group product to show that
\begin{equation}
    F_t = f\circ f_t=f\circ (\mathrm{id} + t\xi + \order{t^2})=(\mathrm{id}+tf_*\xi)\circ f+\order{t^2},
\end{equation}
and
\begin{equation}\label{eq: Ht}
    H_t = h f_* h_t = h(1 + t f_* J)+ \order{t^2}.
\end{equation}
Therefore
\begin{equation}
\label{eq: Kt}
    K_t\equiv F_{t*}(f^* h^{-1})=(F_t\circ f^{-1})_* h^{-1}=h^{-1}\qty(1 + t (\ldv{f_*\xi}h)h^{-1}) + \order{t^2}.
\end{equation}
Note that this can be equivalently written as
\begin{equation}
    K_t  = h^{-1}(1 + t \iota_{f_*\xi} R)+ \order{t^2},
\end{equation}
where we denoted by $R\equiv R[h] = \dd h h^{-1}$ the right-invariant Maurer--Cartan form. The Mickelsson cocycle is then
\begin{equation}
    \COC_{\KM}(H_t,K_t) = \frac{1}{8\pi}\int_D \Tr[H_t^{-1}\dd H_t\wedge \dd K_t K^{-1}_t].
\end{equation}
Using equations~\eqref{eq: Ht} and~\eqref{eq: Kt}, we can evaluate the two elements separately
\begin{align}
    H_t^{-1} \dd H_t &= L  + t (\dd(f_*J)+ \qty[L,f_*J]) + \order{t^2},\\
    \dd K_t K_t^{-1} &= - L + t h^{-1}\dd Z h + \order{t^2},
\end{align}
where we introduced the left-invariant Maurer--Cartan form $L\equiv L[h]= h^{-1}\dd h$ and denoted $Z\equiv \iota_{f_*\xi}R$. Putting these together, we find
\begin{equation}
    \dv{t}\COC_{\KM}(H_t,K_t)\eval_{t=0} = \frac{1}{8\pi}\int_D \Tr[L\wedge\qty(\dd (f_* J) + \qty[L,f_* J]+ h^{-1}\dd Z h)].
\end{equation}
Summing this and~\eqref{eq: Souriau computation first term} we obtain
\begin{Align}
\label{eq: Souriau computation}
\dv{t}&\qty(\COC^{\Diff(S)}_{\KM}(g,e^{tX}) + \COC^{\Diff(S)}_{\KM}(ge^{tX},g^{-1}))\eval_{t=0}=\\ &\quad= \frac{1}{8\pi}\int_D \Tr[L\wedge\qty(2\dd (f_* J) + \qty[L,f_* J]+ h^{-1}\dd Z h)]=\\
&\quad= \frac{1}{8\pi}\int_D \Tr[2 L\wedge \dd (f_* J) +2 (f_* J) L \wedge L +R\wedge \dd Z]=\\
&\quad = -\frac{1}{4\pi}\oint\Tr[(f_* J)L] - \frac{1}{8\pi}\oint\Tr[R Z]+\frac{1}{8\pi}\int_D\Tr[\dd R Z],
\end{Align}
where we used the cyclicity of the trace and the Maurer--Cartan equation for $L$ (namely $\dd L+L\wedge L=0$). Notice that the bulk term drops using the equation for $R$ (namely $\dd R-R\wedge R=0$) because, thanks to the trace cyclicity, one can reduce it to the contraction of a three-form
\begin{equation}
    \Tr[(R\wedge R)\, \iota_{f_*\xi }R]=\frac{1}{3}\iota_{f_*\xi }\Tr[R\wedge R\wedge R]
\end{equation}
which necessarily vanishes on $D$.

The first term in~\eqref{eq: Souriau computation} can be written as
\begin{equation}
    \oint \Tr[(f_* J)L]=\oint \Tr[L[f^* h] J]
\end{equation}
which is the natural pairing in $L\spl{2}$. Moreover, we can permute $h^{-1}$ factors in the second boundary term to rewrite it as
\begin{equation}
\label{eq: let's finish}
    \oint \Tr[RZ]=\oint \Tr[R[h]\iota_{f_*\xi} R[h]]=\oint \iota_\xi\Tr[L[f^* h]\otimes L[f^* h]],
\end{equation}
The final result can then be written as
\begin{equation}
-\frac{1}{4\pi}\oint \Tr[L[f^*h]J] - \frac{1}{8\pi}\oint \iota_\xi \Tr[L[f^* h]\otimes L[f^* h]]=-\frac{1}{24\pi}\pair{\mathsf S_{\KM}[g]}{(\xi,J,b)},
\end{equation}
using the definition~\eqref{eq: sourieaudefinition}, from which we extract the Kac--Moody Souriau cocycle
\begin{equation}
    \mathsf{S}_{\KM}[f,h,\alpha] = \qty(3\Tr[L[f^*h]\otimes L[f^*h]],6\Tr[L[f^*h],\mdot],0).
\end{equation}

\section{Details on irreducibility}
\label{app:irreducibility}

In this appendix, we provide the details of the irreducibility arguments outlined in the main text. We first study the commutant of the full representation and its reduction to the commutant of the little group representation. We then discuss the ergodicity properties required on the momentum and Casimir density orbits.

\subsection{Decomposability with respect to the momentum orbit}\label{app:C1}

As in the main-text argument (cfr.~around Eq.~\eqref{eq: T acts as multiplication}), we let $T$ be a bounded operator that commutes with all the operator of the $\ECS_3$ representation.
In particular, if we introduce for $\alpha\in L\RR^2$, the representation operator
\begin{equation}
    [M_\alpha\Psi](p,\nu):= \chi_p(\alpha)\Psi(p,\nu)= e^{i\oint\pair{p}{\alpha}}\,\Psi(p,\nu),
\end{equation}
we have
\begin{equation}
    T M_\alpha = M_\alpha T, \quad \forall\alpha\in L\RR^2\,.
\end{equation}
We now consider the $\sigma$-algebra generated by the characters
\begin{equation}
    \Sigma_\chi
    \defeq
    \sigma\left(
        \chi_\alpha^{-1}(\mathcal A)
        \,\middle|\,
        \alpha\in L\RR^2,\,
        \mathcal A\in\mathcal B(\U{1})
    \right).
\end{equation}
and claim that
\begin{equation}\label{eq:app-borel-momentum}
    \Sigma_\chi=\mathcal B(\mathcal O_{p_\star}).
\end{equation}
The inclusion $\Sigma_\chi\subseteq\mathcal B(\mathcal O_{p_\star})$ is immediate, since every $\chi_\alpha$ is continuous and hence Borel measurable. To prove the reverse inclusion $\Sigma_\chi\supseteq\mathcal B(\mathcal O_{p_\star})$, we expand $p\in\mathcal O_{p_\star}$ in Fourier modes
\begin{equation}
    p_a(\x)=\sum_{n\in\mathbb Z}p_{a,n}e^{in\x},\qquad a=0,1.
\end{equation}
The Borel $\sigma$-algebra inherited from the
$C^\infty(S^1,\RR^2)$ Fr\'echet topology is generated by the Fourier coordinate maps $\Re p_{a,n}$ and $\Im p_{a,n}$. It is easy to see that each of these coordinates can be obtained from a pairing with $L\RR^2$, explicitly
\begin{equation}
    \Re p_{\alpha, n}=\frac{1}{2\pi}\oint \pair{p}{e_a \cos(n\x)},\quad \Im p_{\alpha, n}=\frac{1}{2\pi}\oint \pair{p}{e_a \sin(n\x)} 
\end{equation}
with $\{e_0,e_1\}$ a basis of $\RR^2$.
It remains to show that one can recover each pairing from the characters themselves: for every $k\in\mathbb N$, $\alpha/k$ is still in $L\RR^2$, hence $\chi_{\alpha/k}(p)$ is measurable in the $\sigma$-algebra $\Sigma_\chi$; on the other hand, 
\begin{equation}
    \oint\pair{p}{\alpha}=\lim_{k\rightarrow\infty}k\,\operatorname{Arg}\qty(e^{i\oint\pair{p}{\alpha}/k}),
\end{equation}
where $\operatorname{Arg}$ is the principal value of the argument. This shows that each pairing --- and therefore each Fourier coordinate --- is also $\Sigma_\chi$-measurable, proving the reverse inclusion.

From~\eqref{eq:app-borel-momentum}, it follows that the operators $M_\alpha$ generate the full algebra of bounded multiplication operators in the $p$ variable. Therefore, by the standard decomposition theorem for operators on vector-valued $L^2$ spaces, $T$ is decomposable with respect to the momentum variable
\begin{equation}\label{eq:app-T-decomposable}
    \bigl([T\Psi](p)\bigr)(\nu)=\bigl(T_p(\Psi(p))\bigr)(\nu)\qquad\mathrm{a.e.},
\end{equation}
where $T_p\in\mathcal B(\mathcal K)$ acts on the target $\nu$-dependent wavefunction $\Psi(p)\in\mathcal K$.\footnote{Strictly speaking, the statement that an operator commuting with
the full multiplication algebra is decomposable as in
Eq.~\eqref{eq:app-T-decomposable} relies on the standard direct-integral
theory for vector-valued $L^2$ spaces, and therefore assumes the existence
of a genuine measure on the momentum orbit. In the present construction,
this statement is rigorous for the regulated, finite-dimensional, measures
introduced in the main text and is assumed to remain valid in their formal
continuum limit.}

\subsection{Reduction to the little group commutant}\label{app:C2}

We here prove the claim we made between Eq.~\eqref{eq: T acts as multiplication} and Eq.~\eqref{eq:Tisconstant}, that the operator $T_p$, defined just above in~\eqref{eq:app-T-decomposable}, belongs to the commutant of the little-group representation.

Let $H=\Diff^{+}\!(S)\ltimes L_e\SL{2}$ denote the group acting non-trivially on the momentum orbit. Given $g\in H$, its induced action can be written in the form
\begin{equation}
    \bigl([U_{g}\Psi](p)\bigr)(\nu)=\bigl(\overcirc{U}_{W(g;p)}\Psi(g^{-1}\lga p)\bigr)(\nu)\,,
\end{equation}
where $\overcirc U$ is the representation of the little group $H_{p_\star}$ and $W(g;p)$ is the Wigner element. For conciseness, we removed the representations labels.
Combining $TU_g=U_gT$ with
\eqref{eq:app-T-decomposable} gives
\begin{equation}\label{eq:app-covariance-Tp}
 T_p \overcirc U_{W(g;p)}=\overcirc U_{W(g;p)} T_{g^{-1}\lga p}
    \qquad
   \mathrm{a.e.}.
\end{equation}
Let us now pick a $k\in H_{p_\star}$. Since the little groups of two elements in the same orbits are conjugated, we can define an element in the little group of $p$ as
\begin{equation}
    g=\sect{s}_p\cdot k \cdot \sect{s}^{-1}_p\,
\end{equation}
where $\sect{s}_p$ is any section that maps $p_\star$ to $p$ --- in the main text, we chose $\sect{s}_p=(\mathrm{id},\sect{h}_p)$. It is easy to see that for this $g$
\begin{equation}
    W(g;p)=k\,.
\end{equation}
Hence, if we apply Eq.~\eqref{eq:app-covariance-Tp} to $g$, we obtain
\begin{equation}
\label{eq: app-commutativity}
    T_p \overcirc U_k=\overcirc U_k T_p\,.
\end{equation}
Since $k$ was arbitrary, this proves that $T_p\in \overcirc U(H_{p_\star})'$.
As argued in the main text, \textit{if} the little group representation is irreducible, we can then use Schur's lemma and Eq.~\eqref{eq: app-commutativity} to conclude that
\begin{equation}\label{eq:app-Tp-scalar}
    T_p
    =
    c(p)\mathbf 1_{\mathcal K}
    \qquad
   \mathrm{a.e.},
\end{equation}
for some measurable function $c(p)$.

\subsection{Ergodicity of the regulated momentum orbit}\label{app:C3}

We establish here the ergodicity property of the measure $\mathcal D p$ under the $L_e\SL{2}$ action. As mentioned in the main text, we are going to show that this holds for the regulated measure at any finite value of the regulator, and postulate that it continues to hold in the limit where we remove the regulator.

Consider, for a given $N\in\NN$, the regulated version of the momentum orbit equipped with a finite-dimensional measure $(\mathcal O^N_{p_\star},\mathcal D_N p)$. Explicitly
\begin{equation}
    \mathcal O_{p_\star}^{N}
    =
    \prod_{i=1}^{N}\RR^2_\circ,\qquad \mathcal D_Np
    =
    \prod_{i=1}^{N}\dd^2p_i.
\end{equation}
The regulated loop group is just a direct product of $N$ independent $\SL{2}$, acting as
\begin{equation}
    (h_1,\ldots,h_N)
    \lga
    (p_1,\ldots,p_N)
    =
    (h_1\lga p_1,\ldots,h_N \lga p_N).
\end{equation}
It is easy to establish that: (i) this action preserves the regulated measure since $\det h_i=1$, $\forall i\le N$, and that (ii) it is transitive because, for arbitrary $p=(p_1,\ldots,p_N)$ and $q=(q_1,\ldots,q_N)$ in $\mathcal O_{p_\star}^{N}$, one can take $h_i=\sect{h}_{q_i}\cdot\sect{h}_{p_i}^{-1}$ for each $i\le N$ and have $h\lga p=q$ by construction.

A transitive, measure-preserving action on a finite-dimensional homogeneous space is ergodic. To see why, let
$\mathcal A\subset\mathcal O_{p_\star}^{N}$ be a measurable invariant subset. Its
indicator function satisfies
\begin{equation}
    \mathbf 1_{\mathcal A}(h^{-1}\rhd p)
    =
    \mathbf 1_{\mathcal A}(p)
    \quad
   \mathrm{a.e.},\quad \forall h\in\SL{2}^N\,.
\end{equation}
Because of transitivity, any invariant measurable function is constant almost everywhere. Given that an indicator function takes only the values zero and one, it follows that either
\begin{equation}
    \int_\mathcal A\mathcal D_Np=0 \qquad \text{or} \qquad \int_{\mathcal O_{p_\star}^N \setminus \mathcal A}\mathcal D_Np=0\,.
\end{equation}
This means that the regulated loop group action is ergodic for every finite $N$. For the continuous case, we postulate that this ergodicity persists when the regulator is removed, in the same formal sense in which the flat functional path-integral measure $\mathcal Dp$ is defined.

Notice that, since the loop subgroup action is already ergodic, it is not necessary to include the regulated diffeomorphism action for this result.

\subsection{Invariant functions under an ergodic action}\label{app:C4}

We give here a proof of the following measure-theoretic statement we used repeatedly in the main text. Let $(X,\mu)$ is a measure space with an ergodic nonsingular action of a group $G$. If a measurable function $f$ on $X$ is such that
\begin{equation}
\label{eq: app G-invariance}
    a(g^{-1}\rhd x)
    =
    a(x)
    \quad
    \text{a.e.},\quad \forall g\in G\,,
\end{equation}
then it is constant almost everywhere.
We prove this statement by contradiction. If $a$ is \textit{not} constant almost everywhere, it means that $\operatorname*{ess\,inf}a$ is strictly smaller than $\operatorname*{ess\,sup}a$.\footnote{The \textit{essential infimum} (\textit{supremum}) of a measurable function $a$ is the largest (smallest) number $c$ ($C$) such that $a(x)\ge c$ ($a(x)\le C$) almost everywhere.} We can then pick $\lambda\in\RR$ such that
\begin{equation}
    \operatorname*{ess\,inf}a<\lambda<\operatorname*{ess\,sup}a\,,
\end{equation}
and define the set
\begin{equation}
    A_\lambda
    =
    \left\{
        x\in X
        \,\middle|\,
        a(x)>\lambda
    \right\}.
\end{equation}
The a.e.~invariance of $a$ implies that $A_\lambda$ is also a.e.~invariant under $G$. On the other hand $\mu(A_\lambda)>0$ by definition of essential supremum, and simultaneously $\mu(X\setminus A_\lambda)>0$ by definition of essential infimum. This means that $A_\lambda$ is an invariant subset with ``intermediate'' measure, contradicting the hyphotesis of ergodicity. 
For a complex-valued function, the same result follows by applying the argument to its real and imaginary parts.

\subsection{Irreducibility of the little group representation}
\label{app:C5}

We establish here the irreducibility of the little-group representation $\overcirc U$ acting on the Hilbert space $\tilde K=L^2(\tilde{\mathcal O}_{\nu_\star},\mu)$, where
\begin{equation}
    \tilde{\mathcal O}_{\nu_\star}
    \cong
    \Diff^1_+\!(S)/\U{1}
    =
    \Diff^1_{+,0}(S)
\end{equation}
is the enlarged Casimir-density orbit obtained by relaxing the smoothness assumption on the diffeomorphism action, and $\mu$ is the corresponding Shavgulidze measure. We use an argument analogous to that in Appendix~\ref{app:C1}. We take $A$ to be a bounded operator that commutes with all operators of the little-group representation. In particular, it commutes with the diagonal operators that represent the $LN$ subgroup
\begin{equation}
    [\overcirc M_x \psi](\nu):=\chi_\nu(x) \psi(\nu)=e^{i\oint\pair{\nu}{x}}\,\psi(\nu).
\end{equation}

We can readily show that the characters $\chi_\nu$ generate the entire Borel $\sigma$-algebra of $\tilde{\mathcal O}_{\nu_\star}$. Indeed, from the pairings
\begin{equation}
    \frac{1}{2\pi}\oint \pair{\nu}{\cos(n\x)(\dd\x)^{-2}},\qquad \frac{1}{2\pi}\oint \pair{\nu}{\sin(n\x)(\dd\x)^{-2}}
\end{equation}
we can recover the Fourier coefficients of $\nu(\x)$ which generate the Borel structure inherited from the $C^0(S)$ topology. Furthermore, since $\frac{x}{k}\in LN$ for every $k\in\mathbb{N}$, the exponentiated characters contain the same measurable information as the linear pairings, exactly as in the momentum-orbit argument of Appendix~\ref{app:C1}. Therefore, it follows that $A$ commutes with the full algebra of bounded multiplication operators on
$\tilde{\mathcal O}_{\nu_\star}$. Since there is no remaining nontrivial fiber at this stage, $A$ must itself be a multiplication operator
\begin{equation}\label{eq:app-S-multiplication}
    [A\psi](\nu)
    =
    a(\nu)\psi(\nu) \quad \text{-a.e.},
\end{equation}
for some measurable function $a$.

The assumption that $A$ commutes with the representation of smooth diffeomorphisms
\begin{equation}
\label{eq: app smooth based diffeo rep}
    [\overcirc U^{\mathrm{RN}}_f\psi](\nu)
    =
    \rho_f(\nu)^{\frac12}
    e^{i m\beta_W(f;\nu)}
    \psi(f^{-1}\rhd\nu),
    \qquad
    f\in\Diff^\infty_{+,0}(S),
\end{equation}
implies, together with~\eqref{eq:app-S-multiplication}, that
\begin{equation}\label{eq:app-a-smooth-invariant}
    a(\nu)
    =
    a(f^{-1}\rhd\nu)
    \quad
    \text{a.e.},\quad f\in\Diff^\infty_{+,0}\!(S).
\end{equation}
In order to apply the result of the previous subsection and claim that $a$ is a constant almost everywhere, we would need the action of $\Diff^\infty_{+,0}(S)$ to be ergodic, but this is not the case. However, as we recall in the main text, a result by Kosyak shows that $(\tilde{\mathcal O}_{\nu_\star},\mu)$ \textit{does} admit an ergodic action under $\Diff^2_{+,0}(S)$.

To take advantage of Kosyak's ergodicity result, we extend the representation~\eqref{eq: app smooth based diffeo rep} of $\Diff^\infty_{+,0}(S)$ to a representation of $\Diff^2_{+,0}(S)$. To do that, we need to assume that the representation is strongly continuous. We then use the fact that $\Diff^\infty_{+,0}(S)$ is dense in $\Diff^2_{+,0}(S)$ in the $C^2$ topology to construct a sequence of smooth diffeomorphisms that approximate any given $f\in \Diff^2_{+,0}(S)$:
\begin{equation}
    f_n\in\Diff^\infty_{+,0}(S),
    \qquad
    f_n
    \xrightarrow[n\rightarrow\infty]{C^2}
    f.
\end{equation}
The strong continuity then guarantees that
\begin{equation}
    \overcirc U^{\mathrm{RN}}_{f_n}\psi
    \longrightarrow
    \overcirc U^{\mathrm{RN}}_{f}\psi\,,
\end{equation}
for every $\psi\in\tilde{\mathcal K}$, independently of the chosen sequence. This defines the operator $\overcirc U^{\mathrm{RN}}_{f}$ for a twice-differentiable but non-smooth diffeomorphism.

Since $A$ is bounded, it also commutes with the operators associated with
$\Diff^2_{+,0}(S)$. Indeed, for any $f\in\Diff^2_{+,0}(S)$, let
$f_n\in\Diff^\infty_{+,0}(S)$ be a sequence converging to $f$ in the
$C^2$ topology. By the assumed strong continuity of the representation,
\begin{equation}
    A \overcirc U^{\mathrm{RN}}_{f}\psi
    =
    A \lim_{n\rightarrow\infty}
    \overcirc U^{\mathrm{RN}}_{f_n}\psi
    =
    \lim_{n\rightarrow\infty}
    A\,\overcirc U^{\mathrm{RN}}_{f_n}\psi
    =
    \lim_{n\rightarrow\infty}
    \overcirc U^{\mathrm{RN}}_{f_n}A\psi
    =
    \overcirc U^{\mathrm{RN}}_{f}A\psi .
\end{equation}
This extends the invariance relation~\eqref{eq:app-a-smooth-invariant} to
\begin{equation}\label{eq:app-a-2-invariant}
    a(\nu)
    =
    a(f^{-1}\rhd\nu)
    \quad
    \text{a.e.},
    \qquad
    \forall f\in\Diff^2_{+,0}\!(S).
\end{equation}

We can now use Kosyak's result that the Shavgulidze measure on
$\Diff^1_{+,0}(S)$ is ergodic under the left action of
$\Diff^2_{+,0}(S)$. Under the identification
\begin{equation}
    \tilde{\mathcal O}_{\nu_\star}
    \cong
    \Diff^1_{+,0}(S),
\end{equation}
equation~\eqref{eq:app-a-2-invariant} therefore states that $a$ is invariant
under an ergodic action. By the result of the previous subsection, any such
measurable invariant function is constant almost everywhere. Hence
\begin{equation}
    A=a\,\mathbf 1_{\tilde{\mathcal K}}
\end{equation}
for some $a\in\CC$. By Schur's lemma, the little-group representation is
therefore irreducible.

\bibliographystyle{JHEP}
\bibliography{3dRepBib}

\end{document}